\documentclass[twocolumn,superscriptaddress,nofootinbib
]{revtex4-1}

\accepted{to JPCA September 22, 2020}

\renewcommand{\thetable}{\arabic{table}}

\makeatletter
\let\@fnsymbol\@arabic
\makeatother

\usepackage{graphicx}
\usepackage{dcolumn}
\usepackage{bm}
\usepackage{mhchem}
\usepackage{longtable}
\usepackage{supertabular}

\usepackage{textpos}
\usepackage{hyperref}
\usepackage{amsmath}
\usepackage{wasysym}

\setcitestyle{numbers,super}

\usepackage[symbol]{footmisc}
\renewcommand{\thefootnote}{\fnsymbol{footnote}}

\newcommand{\beginsupplement}{%
        \setcounter{table}{0}
        \renewcommand{\thetable}{S\arabic{table}}%
        \setcounter{figure}{0}
        \renewcommand{\thefigure}{S\arabic{figure}}%
        \setcounter{equation}{0}
        \renewcommand{\theequation}{S\arabic{equation}}%
     }

\makeatletter
\let\cat@comma@active\@empty
\makeatother

\begin{document}

\title{CRAHCN-O: A Consistent Reduced Atmospheric Hybrid Chemical Network Oxygen Extension for Hydrogen Cyanide and Formaldehyde Chemistry in CO$_2$-, N$_2$-, H$_2$O-, CH$_4$-, and H$_2$-Dominated Atmospheres}

\author{Ben K. D. Pearce}
\email[Corresponding author:]{pearcbe@mcmaster.ca}
\affiliation{Origins Institute and Department of Physics and Astronomy, McMaster University, ABB 241, 1280 Main St, Hamilton, ON, L8S 4M1, Canada}
\author{Paul W. Ayers}
\affiliation{Origins Institute and Department of Chemistry and Chemical Biology, McMaster University, ABB 156, 1280 Main St, Hamilton, ON, L8S 4M1, Canada}
\author{Ralph E. Pudritz}
\affiliation{Origins Institute and Department of Physics and Astronomy, McMaster University, ABB 241, 1280 Main St, Hamilton, ON, L8S 4M1, Canada}

\renewcommand*{\thefootnote}{\arabic{footnote}}

\begin{abstract}
{\bf Abstract:} Hydrogen cyanide (HCN) and formaldehyde (\ce{H2CO}) are key precursors to biomolecules such as nucleobases and amino acids in planetary atmospheres; However, many reactions which produce and destroy these species in atmospheres containing \ce{CO2} and \ce{H2O} are still missing from the literature. We use a quantum chemistry approach to find these missing reactions and calculate their rate coefficients using canonical variational transition state theory and Rice--Ramsperger--Kassel--Marcus/master equation theory at the BHandHLYP/aug-cc-pVDZ level of theory. We calculate the rate coefficients for 126 total reactions, and validate our calculations by comparing with experimental data in the 39\% of available cases. Our calculated rate coefficients are most frequently within an factor of 2 of experimental values, and generally always within an order of magnitude of these values. We discover 45 previously unknown reactions, and identify 6 from this list that are most likely to dominate \ce{H2CO} and HCN production and destruction in planetary atmospheres. We highlight \ce{^1O + CH3 -> H2CO + H} as a new key source, and \ce{H2CO + ^1O -> HCO + OH} as a new key sink, for \ce{H2CO} in upper planetary atmospheres.
In this effort, we develop an oxygen extension to our consistent reduced atmospheric hybrid chemical network (CRAHCN-O), building off our previously developed network for HCN production in \ce{N2}-, \ce{CH4}- and \ce{H2}-dominated atmospheres (CRAHCN). This extension can be used to simulate both HCN and \ce{H2CO} production in atmospheres dominated by any of \ce{CO2}, \ce{N2}, \ce{H2O}, \ce{CH4}, and \ce{H2}.
\end{abstract}

\maketitle


\section*{Introduction}

Hydrogen cyanide (HCN) and formaldehyde (\ce{H2CO}) are key precursors to various biomolecules required for the origin of life. The four nucleobases in RNA, i.e., adenine, guanine, cytosine and uracil, form in aqueous solutions containing one or both of these reactants \citep{1961Natur.191.1193O,2008OLEB...38..383L,2019AA...626A..52F}. Ribose, which pairs with phosphate to make up the backbone of RNA, forms from the oligomerization of \ce{H2CO} \citep{Butlerow1861,Breslow1959}. Amino acids form via Strecker synthesis, which includes both HCN and an aldehyde (\ce{H2CO} for glycine) as reactants \citep{Strecker1854,Miller_VanTrump1981}.

Given their substantial role in producing biomolecules, HCN and \ce{H2CO} may be distinguishing atmospheric features of what we call \emph{biogenic worlds}. These are worlds capable of producing key biomolecules rather than requiring they be delivered (e.g., by meteorites). It is presently unknown whether the early Earth was biogenic.

The redox state of the oldest minerals on the planet suggests the early Earth atmosphere was composed of ``weakly reducing'' gases, i.e., \ce{CO2}, \ce{N2}, and \ce{H2O}, with relatively smaller amounts of \ce{CH4}, \ce{CO}, and  \ce{H2} \citep{2007AsNow..22e..76R,Reference119}. These atmospheric species are broken up into reactive radicals by UV radiation, lightning, and/or galactic cosmic rays (GCRs), which allows disequilibrium chemistry and the production of HCN and \ce{H2CO} to occur \citep{2007AsNow..22e..76R,Pinto1980}. The following pathways are possible from the dissociation of these ``weakly reducing'' species\citep{Schmidt2013,2017ApJ...850...48S,Reference586,Engel1992,1967ApJ...149L..29S}: 

\begin{equation}
\ce{CO2 + $h \nu$ -> CO + ^3O}
\end{equation}
\begin{equation}
\ce{-> CO + ^1O}
\end{equation}


\begin{equation}
\ce{N2 + $h \nu$ -> ^4N + ^2N}
\end{equation}

\begin{equation}
\ce{CH4 + $h \nu$ -> CH3 + H}
\end{equation}
\begin{equation}
\ce{-> ^3CH2 + 2H}
\end{equation}
\begin{equation}
\ce{-> ^1CH2 + H2}
\end{equation}
\begin{equation}
\ce{-> CH + H2 + H}
\end{equation}

\begin{equation}
\ce{H2O + $h \nu$ -> OH + H}
\end{equation}

\begin{equation}
\ce{H2 + $h \nu$ -> 2H}
\end{equation}
where the superscripts, $^1$, $^2$, $^3$, and $^4$ refer to the singlet, doublet, triplet and quartet electronic spin states.

One way to better understand the biogenicity of the early Earth, is to use chemical kinetic models to simulate the production of HCN and \ce{H2CO} in plausible early Earth atmospheres. Atmospheric simulations of these species for primitive Earth conditions have been performed in the past \citep{Pinto1980,Reference591,2011EPSL.308..417T}, which make use of collections of reaction rate coefficients typically gathered from various sources the literature (e.g. experiment, theoretical simulations, thermodynamics, similar reactions).

The literature, however, is still missing several reactions between the radicals produced in \ce{CO2}-, \ce{N2}-, \ce{H2O}-, \ce{CH4}-, and \ce{H2}-dominated atmospheres, and these reactions may be crucial to understanding HCN and \ce{H2CO} chemistry in early Earth and other terrestrial environments. The largest gap in rate coefficient data is for reactions involving electronically excited species, e.g. \ce{^1O}, \ce{^2N}, and \ce{^1CH2}, which are directly produced from the dissociation of \ce{CO2}, \ce{N2}, and \ce{CH4}, respectively.

In \citet{Pearce2020a} and \citet{Reference598}, we developed an accurate and feasible method making use of computational quantum chemistry coupled with canonical variational transition state theory (CVT) \citep{Reference534} and Rice--Ramsperger--Kassel--Marcus/master equation (RRKM/ME) theory \citep{Reference2058} to calculate a large network of reaction rate coefficients for one-, two- and three-body reactions. We first used this method to explore the entire field of possible reactions for a list of primary species in \ce{N2}-, \ce{CH4}-, and \ce{H2}-dominated atmospheres, and uncovered 48 previously unknown reactions; many of which were based on excited species such as \ce{^2N} and \ce{^1CH2}. We then built an initial reduced network of 104 reactions based on this exploratory study, and used it to simulate HCN production in Titan's atmosphere \citep{Pearce2020a}. This approach provided us with a more complete picture of HCN chemistry on Titan, as one of our newly discovered reactions was found to be one of the four dominant channels to HCN production on Titan \citep{Pearce2020a}.

In this work, we use the same theoretical approach to expand upon our initial network, by exploring and calculating all the potential reactions between three key oxygen species present on the early Earth (\ce{CO2}, \ce{H2O}, \ce{H2CO}), their dissociation radicals (\ce{CO}, \ce{^3O}, \ce{^1O}, \ce{OH}, and \ce{HCO}), and all the non-oxygen primary species in our network (see Table~\ref{Table1} for the list of primary species). In this effort, we discover 45 brand new reactions, which are mainly based on \ce{HCO}, \ce{H2CN}, \ce{^1O}, \ce{^2N}, \ce{^1CH2}, and \ce{CH}. We calculate the rate coefficients for a total of 126 reactions, and validate our calculations by comparing with experimental data in the 39\% of available cases. 

Finally, we build the consistent reduced atmospheric hybrid chemical network oxygen extension (CRAHCN-O), composed of experimental rate coefficients when available, and our calculated values otherwise. CRAHCN-O is the amalgamation of the network developed in \citet{Pearce2020a}, and the oxygen reactions explored in this work. This network can be used to accurately simulate HCN and \ce{H2CO} production in \ce{CO2}-, \ce{N2}-, \ce{H2O}-, \ce{CH4}-, and \ce{H2}-dominated atmospheres. 

The paper is outlined as follows: In the Methods section, we detail the theoretical and computational approach we use to explore reactions and calculate their rate coefficients. In the Results section, we describe the results of our rate coefficient calculations, including their agreement with any available experiments. We also discuss the limitations of our theoretical approach. In the Discussion section, we highlight 6 new reactions from this work which are potentially key production and destruction pathways to \ce{H2CO} and \ce{HCN} in planetary atmospheres. We also summarize CRAHCN-O and describe how it can be used for other atmospheric models. Finally, in the Conclusions section, we summarize the main conclusions from this work.

The Supporting Information (SI) contains two tables summarizing the new CRAHCN-O rate coefficient data (the non-oxygen reaction data can be found in \citet{Pearce2020a}), any experimental rate coefficient data for reactions calculated in this work, the Lennard-Jones parameters used for three-body reaction rate coefficient calculations, a breakdown of some of the non-standard reaction calculations, and the quantum chemistry data used in our calculations.


\begin{table*}[t]
\centering
\caption{List of primary molecular species involved in this study and their spin states. Reactions strictly between non-oxygen species (below center line) were explored in \citet{Pearce2020a} and \citet{Reference598}. Reactions involving the oxygen species (above center line) are new to this study. \label{Table1}} 
\begin{tabular}{ccc}
\\
\multicolumn{1}{c}{Species} & 
\multicolumn{1}{c}{\hspace{0.3cm}Spin state} & 
\multicolumn{1}{c}{\hspace{0.3cm}Ground/Excited state}\\ \hline \\[-2mm]
CO$_2$ & singlet & ground \\
H$_2$CO & singlet & ground \\
HCO & doublet & ground \\
CO & singlet & ground \\
H$_2$O & singlet & ground \\
 OH & doublet & ground \\
$^3$O  & triplet & ground \\
$^1$O  & singlet & excited \\
\hline \vspace{-2mm} \\
H$_2$CN & doublet & ground \\
HCN & singlet & ground \\
CN & doublet & ground \\
N$_2$ & singlet & ground \\
NH & triplet & ground \\
$^2$N & doublet & excited \\
$^4$N & quartet & ground \\
CH$_4$ & singlet & ground \\
CH$_3$ & doublet & ground \\
$^1$CH$_2$ & singlet & excited \\
$^3$CH$_2$ & triplet & ground \\
CH & doublet & ground \\
H$_2$ & singlet & ground \\
H & doublet & ground \\
\hline
\end{tabular}
\end{table*}

\section*{Methods}\label{methods}

There are four phases to this work: First we explore all the potential reactions between eight oxygen species (\ce{CO2}, \ce{CO}, \ce{^3O}, \ce{^1O}, \ce{H2O}, \ce{OH}, \ce{H2CO}, and \ce{HCO}) and the primary species in Table~\ref{Table1}. These species are the the dominant sources of oxygen in the early Earth atmosphere (\ce{CO2} and \ce{H2O}), a key biomolecule precursor (\ce{H2CO}) and their dissociation radicals. In this process, we characterize 81 known reactions and discover 45 previously unknown reactions. Second, we calculate the rate coefficients for every reaction that we find at 298 K, and validate the calculations by comparing to experimental data when available (in 39\% of cases). Third, we calculate the temperature dependencies for the reactions that have no experimental measurements and have barriers (i.e. strong temperature dependencies from 50--400 K). Last, we gather the experimental and theoretical rate coefficients into the consistent reduced atmospheric hybrid chemical network oxygen extension (CRAHCN-O), which contains experimental values when available, and our calculated rate coefficients otherwise.

\subsection*{Computational Quantum Method and Basis Set}

All exploration and rate coefficient calculations are performed with the Becke-Half-and-Half-Lee-Yang-Parr\footnote{This hybrid functional uses 50\% Hartree-Fock (HF) and 50\% density functional theory (DFT) for the exchange energy calculation, offering a compromise between HF, which tends to overestimate energy barriers, and DFT, which tends to underestimate energy barriers.} (BHandHLYP) density functional and the augmented correlation-consistent polarized valence double-$\zeta$ (aug-cc-pVDZ) basis set \citep{Reference594,Reference595,Dunning1989,KendallDunning1992,WoonDunning1993}. 

We have four key reasons for choosing this level of theory to perform our calculations: 

1) We have benchmarked BHandHLYP/aug-cc-pVDZ rate coefficient  calculations by comparing with experimental values in the past, and this method most frequently provides the best accuracy with respect to agreement with experimental values in comparison with other widely used, computationally cost effective methods. 

In \citet{Pearce2020a}, we compared the accuracy of 3 methods for calculating 12 reaction rate coefficients. We found BHandHLYP/aug-cc-pVDZ rate coefficient calculations give the best, or equal to the best agreement with experiment in 8 out of 12 cases. This is compared to $\omega$B97XD/aug-cc-pVDZ and CCSD/aug-cc-pVTZ, which gave the best, or equal to the best agreement with experiment in 7 out of 12 and 6 out of 12 cases, respectively \citep{Pearce2020a}. In another method-comparison study on a single reaction between BHandHLYP, CCSD, CAM-B3LYP, M06-2x, B3LYP and HF, all with the aug-cc-pVDZ basis set, we found that only BHandHLYP and CAM-B3LYP provide rate coefficients within the experimental range \citep{Reference598}. 

2) BHandHLYP/aug-cc-pVDZ calculations paired with CVT and RRKM/ME theory typically compute rate coefficients within a factor of two of experimental values, and all calculations generally fall within an order of magnitude of experimental values. This accuracy is consistent with typical uncertainties assigned in large-scale experimental data evaluations \citep{Reference451,Reference509}.

For examples in our network, \citet{Reference451} assign uncertainties of 2--3 to \ce{HCO + HCO -> H2CO + CO} and \ce{^3O + CH -> CO + H} and order-of-magnitude uncertainties to \ce{CO2 + CH -> products}, \ce{H2O + CH -> products}, and \ce{H2CO + CH -> products}.

3) BHandHLYP/aug-cc-pVDZ calculations are computationally cost effective, and therefore feasible for a large scale exploratory study such as ours.

We have also shown in previous work for 12 rate coefficients, that increasing the basis set to the more computationally expensive aug-cc-pVTZ level does not increase the accuracy of our calculations with respect to agreement with experimental values \citep{Pearce2020a}.

4) Finally, using the BHandHLYP/aug-cc-pVDZ level of theory for all the calculations in this oxygen extension allows us to maintain consistency with the calculations in the original network (CRAHCN \citep{Pearce2020a}).

\subsection*{Reaction Exploration}

Using the Gaussian 09 software package \citep{g09}, we perform a thorough search for reactions between eight oxygen species (\ce{CO2}, \ce{CO}, \ce{^3O}, \ce{^1O}, \ce{H2O}, \ce{OH}, \ce{H2CO}, and \ce{HCO}) and the 22 primary species in this study (see Table~\ref{Table1}). The procedure below is carried out for 8$\times$22 = 176 pairs of species.

Using the Avogadro molecular visualization software \citep{avogadro,avogadro2}, we placed each species at a handful of different distances and orientations form its reaction partner. We use a bit of chemical intuition when determining the distance between the species, e.g., abstraction reactions in our network tend to occur at short separations (1--2 $\AA$), whereas addition reactions tend to be longer range (2--6 $\AA$). 

We then copy the geometries into Gaussian input files, and use the `opt=modredundant' option to freeze the bond distances between one atom of each species. We run the Gaussian simulations with vibrational analyses to allow us to identify whether points along the MEP were found. A point along a MEP is identified by a single negative frequency that oscillates in the direction of the reaction. We run multiple simulations to look for possible abstraction, addition, and bond insertion reactions. For reactions that form a single product, we continue the exploration of that product by searching for efficient decay and/or isomerization pathways. In many cases, we find the product efficiently decays into other products, sometimes after one or more isomerizations.

For cases where our above approach fails to find a MEP, we have developed a Python program that can be used to perform a more thorough scan of the potential energy surface. This program takes two species geometries as input, selects, e.g., 10 random separations and orientations for those species, and runs those Gaussian simulations in parallel. This program is especially useful for MEPs that turn out to be not strictly intuitive (e.g. \ce{OH + ^1CH2}).

Once we find a point along a MEP, we then characterize the reaction path by doing a coarse-grain scan backwards and forwards from the identified point in intervals of 0.1$\AA$. We then plot the Gibbs free energies of these optimized points along the reaction path, and analyze the points using Avogadro to find the rough location(s) of the transition state(s). In several cases we find more than one transition state along a reaction path, with one or more stable structures between the reactants and the products.

\subsection*{Rate coefficient calculations}

\subsubsection*{One- and Two-body Reactions}

We calculate one- and two-body reaction rate coefficients using canonical variational transition state theory (CVT). This is a statistical mechanics approach which makes use of the canonical ensemble. This method can be used to calculate rate coefficients for reactions with and without energy barriers \citep{Reference534}.

CVT can be explained as follows. There is a point that is far enough along the minimum energy reaction path (MEP), that the reactants that cross over this point are unlikely to cross back. This point is defined as the location where the generalized transition state (GT) rate coefficient is at its smallest value, therefore providing best dynamical bottleneck \cite{Reference534}. This is expressed as: \citep{Reference535}

\begin{equation}\label{CVT}
k_{CVT}(T,s) = \min_s \left\lbrace k_{GT}(T,s) \right\rbrace.
\end{equation}
where $k_{GT}(T,s)$ is the generalized transition state theory rate coefficient, $T$ is the temperature, and $s$ is a point along the MEP (e.g. bond distance).

To find the location along the MEP where the rate coefficient is at a minimum, we use the maximum Gibbs free energy criterion \citep{Reference535,Reference538}. It can be seen from the quasi-thermodynamic equation of transition-state theory that the maximum value for $\Delta G_{GT}(T,s)$ corresponds to a minimum value for $k_{GT}(T,s)$.

\begin{equation}\label{GibbsCrit}
k_{GT}(T,s) = \frac{k_B T}{h} K^0 e^{-\Delta G_{GT}(T,s)/RT},
\end{equation}
where $K^0$ is the reaction quotient under standard state conditions (i.e. 1 cm$^3$ for second-order reactions, 1 cm$^6$ for third-order reactions), and $\Delta G_{GT}(T,s)$ is the difference in the Gibbs free energy between transition state and reactants (kJ mol$^{-1}$).

This method offers a compromise of energetic and entropic effects, as $\Delta G$ contains both enthalpy and entropy \citep{Reference535,Reference538}. To obtain a similar accuracy for all calculations, we refine our coarse grain scans near the Gibbs maxima to a precision of 0.01 $\AA$.

The generalized transition state theory rate coefficient, neglecting effects due to tunneling, can be calculated with the equation\citep{Reference530,Reference535}

\begin{equation}\label{Eyring}
k_{GT}(T,s) = \sigma \frac{k_B T}{h} \frac{Q^{\ddagger}(T,s)}{\prod_{i=1}^{N} Q_i^{n_i}(T)} e^{-E_0(s)/RT}.
\end{equation}
where $\sigma$ is the reaction path multiplicity, {\it k$_B$} is the Boltzmann constant (1.38$\times$10$^{-23}$ J K$^{-1}$), {\it T} is temperature (K), {\it h} is the Planck constant (6.63$\times$10$^{-34}$ J$\cdot$s), {\it Q$^{\ddagger}$} is the partition function of the transition state per unit volume (cm$^{-3}$), with its zero of energy at the saddle point, {\it Q$_i$} is the partition function of species $i$ per unit volume, with its zero of energy at the equilibrium position of species $i$, $n_i$ is the stoichiometric coefficient of species $i$, $N$ is the number of reactant species, {\it E$_0$} is the difference in zero-point energies between the generalized transition state and the reactants (kJ mol$^{-1}$) (0 for barrierless reactions), and {\it R} is the gas constant (8.314$\times$10$^{-3}$ kJ K$^{-1}$ mol$^{-1}$).

The partition functions per unit volume have four components and are gathered from the Gaussian output files,

\begin{equation}
Q = \frac{q_t}{V} q_e q_v q_r.
\end{equation}
where V is the volume (cm$^{-3}$) and the $t$, $e$, $v$, and $r$ subscripts stand for translational, electronic, vibrational, and rotational, respectively.

In some cases, there are multiple steps (i.e. transition states) to a single reaction, and we must use mechanistic modeling in order to determine the steady-state solution of the overall rate equation. We place an example of a mechanistic model in Case Study 9 in the SI.

\subsubsection*{Three-body reactions}

In the cases where two reactants form a single product, a colliding third body is required to remove excess vibrational energy from the product to prevent it from dissociating \citep{Vallance_Book}. This is expressed as,

\begin{equation}\label{third-order}
\ce{A + B -> C$_{(\nu)}$}
\end{equation}
\begin{equation}
C_{(\nu)} \xrightarrow{+ M} C.
\end{equation}

The rate coefficient for these three-body reactions is expressed as\citep{Reference2064}:

\begin{equation}
k([M]) = \frac{k_0 [M] / k_{\infty}}{1 + k_0 [M] / k_{\infty}} k_{\infty}
\end{equation}
where $k_0$ is the third-order low-pressure limit rate coefficient (cm$^{6}$s$^{-1}$), [M] is the number density of the colliding third body, and $k_{\infty}$ is the second-order high-pressure limit rate coefficient (cm$^{3}$s$^{-1}$).

The high-pressure limit rate coefficients are equivalent to the two-body reaction rate coefficients (i.e., \ce{A + B -> C}), and can be calculated using CVT as above. We make use of the ktools code of the Multiwell Program Suite for the high pressure limit rate coefficient calculations \citep{multiwell2020,Reference2054,Reference2055}.

The low-pressure limit rate coefficients, on the other hand, require information about the collisional third body for their calculation. To calculate these values, we use the Multiwell Master Equation (ME) code, which employs RRKM theory. The ME contains the probabilities that the vibrationally excited product will be stabilized by a colliding third body \citep{Reference2065}. Multiwell employs Monte Carlo sampling of the ME to build up a statistical average for the two outcomes of the reaction (i.e., destabilize back into reactants, or stabilize the product).

With the output from these stochastic trials, we calculate the low-pressure limit rate coefficient with the following equation \citep{Reference2054,Reference2061}:

\begin{equation}
k_0([M]) = \frac{k_{\infty} f_{prod}}{[M]}
\end{equation}
where $k_{\infty}$ is the high-pressure limit rate coefficient, $f_{prod}$ is the fractional yield of the collisionally stabilized product, and [M] is the simulation number density (cm$^{-3}$), which we lower until $k_0$ converges.

We simulate three-body reactions using three different colliding bodies, corresponding to potential dominant species in the early Earth atmosphere (\ce{N2}, \ce{CO2}, and \ce{H2}). The energy transfer was treated with a standard exponential-down model with $<\Delta E>_{down}$ = 0.8 T K$^{-1}$ cm$^{-1}$ \citep{Reference2059,Reference2060}. The Lennard-Jones parameters for the bath gases and all the products were taken from the literature \citep{Reference2057,Reference2056,jetsurf2.0} and can be found in Table S4.

In some cases, when two reactants come together to form a single product, the vibrationally excited product preferably decays along a different channel into something other than the original reactants (e.g. \ce{^1O + H2 -> H2O$_{(\nu)}$* -> OH + H}). In these cases, we also include the second-order reactions to these favourable decay pathways in our network. We verify the preferred decay pathways of vibrationally excited molecules by looking at previous experimental studies.

\subsubsection*{Temperature dependencies}

For the one- and two-body reactions in this study with barriers, and no experimental measurements, we calculate temperature dependencies for the rate coefficients in the 50--400 K range. Barrierless reaction rate coefficients do not typically vary by more than a factor of $\sim$3 between 50 and 400 K \citep{Reference587,Reference495,Reference481,Reference577,Reference572}. To obtain temperature dependencies, we calculate the rate coefficients at 50, 100, 200, 298.15, and 400 K and fit the results to the modified Arrhenius expression

\begin{equation}
k(T) = \alpha \left(\frac{T}{300}\right)^{\beta} e^{-\gamma/T},
\end{equation}
where $k(T)$ is the temperature-dependent second-order rate coefficient (cm$^{3}$s$^{-1}$), $\alpha$, $\beta$, and $\gamma$ are fit parameters, and $T$ is temperature (in K).

\section*{Results}\label{results}

\subsection*{Comparison with Experiments}

In Table~\ref{Results1} we display the three-body high- and low-pressure limit calculated rate coefficients at 298 K. Out of these 31 reactions, 12 have experimentally measured high-pressure limit rate coefficients. For the low-pressure limit rate coefficients, 9 of the 31 reactions have experimental measurements; However, the bath gases used in the low-pressure experiments often differ from the colliding third bodies in our calculations (i.e. \ce{N2}, \ce{CO2}, and \ce{H2}). When using several different bath gases, low-pressure limit rate coefficients tend to range by $\sim$ an order of magnitude \citep{Reference1160,Reference1161,Reference1110,Reference451}.

\setlength\LTcapwidth{\textwidth}
\begin{longtable*}{clcccccc}
\caption{Lindemann coefficients for the three body reactions in this paper, calculated at 298 K, and valid within the 50--400 K temperature range. $k_{\infty}$ and $k_0$ are the third-order rate coefficients in the high and low pressure limits, with units cm$^{3}$s$^{-1}$ and cm$^{6}$s$^{-1}$, respectively. These values are for usage in the pressure-dependent rate coefficient equation $k$ = $\frac{k_0 [M] / k_{\infty}}{1 + k_0 [M] / k_{\infty}} k_{\infty}$. Calculations are performed at the BHandHLYP/aug-cc-pVDZ level of theory. Low-pressure limit rate coefficients are calculated for three different bath gases (\ce{N2}, \ce{CO2}, and \ce{H2}). Reactions with rate coefficients slower than k$_{\infty}$ = 10$^{-13}$ cm$^{3}$s$^{-1}$ are not included in this network. The error factor is the multiplicative or divisional factor from the nearest experimental or suggested value. \label{Results1}} \\
No. & Reaction equation & k$_\infty$(298) calc. & k$_\infty$(298) exp. & Error$_\infty$ & k$_0$(298) calc. & k$_0$(298) exp. & Error$_0$ \\ \hline \\[-2mm]
*1. & \ce{CO2 + ^1O + M -> CO3 + M} & 3.8$\times$10$^{-11}$ & 0.1--23$\times$10$^{-11}$ & 1 &  (M=\ce{N2}) 3.0$\times$10$^{-29}$ & & \\
& & &  & & (\ce{CO2}) 3.1$\times$10$^{-29}$ & & \\
& & &  & & (\ce{H2}) 6.7$\times$10$^{-29}$ & & \\
& & &  & & & & \\
*2. & \ce{HCO + ^2N + M -> } & 2.0$\times$10$^{-11}$ & & & (\ce{N2}) 5.0$\times$10$^{-30}$ & & \\
 & \ce{HCON* + M* -> HCNO + M} & & & & (\ce{CO2}) 5.6$\times$10$^{-30}$ & & \\
& & &  & & (\ce{H2}) 9.7$\times$10$^{-30}$ & & \\
& & &  & & & & \\
*3. & \ce{HCO + CH3 + M -> CH3CHO + M} & 5.7$\times$10$^{-12}$ & 6.3--44$\times$10$^{-12}$ & 1 & (\ce{N2}) 5.3$\times$10$^{-27}$ & & \\
& & &  & & (\ce{CO2}) 6.4$\times$10$^{-27}$ & & \\
& & &  & & (\ce{H2})  1.2$\times$10$^{-27}$ & & \\
& & &  & & & & \\
4. & \ce{HCO + H + M -> H2CO + M} & 4.9$\times$10$^{-11}$ &  &  & (\ce{N2}) 7.4$\times$10$^{-30}$ & & \\
& & &  & & (\ce{CO2}) 9.5$\times$10$^{-30}$ & & \\
& & &  & & (\ce{H2}) 1.4$\times$10$^{-29}$ & & \\
& & &  & & & & \\
*5. & \ce{CO + CN + M -> NCCO + M} & 6.0$\times$10$^{-12}$ & & & (\ce{N2}) 6.2$\times$10$^{-31}$ & & \\
& & &  & & (\ce{CO2}) 6.8$\times$10$^{-31}$ & & \\
& & &  & & (\ce{H2}) 1.3$\times$10$^{-30}$ & & \\
& & &  & & & & \\
6. & \ce{CO + ^1O + M -> CO2 + M} & 2.8$\times$10$^{-11}$ & 0.3--7$\times$10$^{-11}$ & 1 & (\ce{N2}) 2.8$\times$10$^{-30}$ & (\ce{CO2}) 2.8$\times$10$^{-29}$ & 10 \\
& & &  & & (\ce{CO2})  3.0$\times$10$^{-30}$ & " & 9 \\
& & &  & & (\ce{H2}) 5.9$\times$10$^{-30}$ & " & 5 \\
& & &  & & & & \\
*7. & \ce{CO + ^1CH2 + M -> CH2CO + M} & 1.3$\times$10$^{-11}$ & & & (\ce{N2}) 1.7$\times$10$^{-28}$ & & \\
& & &  & & (\ce{CO2}) 1.9$\times$10$^{-28}$ & & \\
& & &  & & (\ce{H2}) 3.3$\times$10$^{-28}$ & & \\
& & &  & & & & \\
8. & \ce{CO + CH + M -> HCCO + M} & 4.6$\times$10$^{-11}$ & 0.5--17$\times$10$^{-11}$ & 1 & (\ce{N2}) 1.2$\times$10$^{-29}$ & \footnotesize (Ar,He) 2.4--4.1$\times$10$^{-30}$ & 3 \\
& & &  &  & (\ce{CO2}) 1.3$\times$10$^{-29}$ & " & 3 \\
& & &  &  & (\ce{H2}) 2.4$\times$10$^{-29}$ & " & 6 \\
& & &  & & & & \\
9. & \ce{CO + H + M -> HCO + M} & 2.7$\times$10$^{-12}$ & & & (\ce{N2}) 1.8$\times$10$^{-33}$ & \footnotesize (Ne,\ce{H2}) 0.5--3.3$\times$10$^{-34}$ & 5 \\
& & &  & & (\ce{CO2}) 2.1$\times$10$^{-33}$ & \footnotesize (CO,\ce{H2}) 0.8--3.3$\times$10$^{-34}$ & 6\\
& & &  & & (\ce{H2}) 3.4$\times$10$^{-33}$ & (\ce{H2}) 0.8--3.3$\times$10$^{-34}$ & 10 \\
& & &  & & & & \\
*10. & \ce{OH + H2CN + M -> H2CNOH + M} & 6.9$\times$10$^{-12}$ & 6.0$\times$10$^{-12}$ & 1 & (\ce{N2}) 6.5$\times$10$^{-30}$ &  &  \\
& & &  & & (\ce{CO2}) 7.4$\times$10$^{-30}$ & & \\
& & &  & & (\ce{H2}) 1.3$\times$10$^{-29}$ & & \\
& & &  & & & & \\
*11. & \ce{OH + CN + M -> HOCN + M} & 1.0$\times$10$^{-12}$ & &  & (\ce{N2}) 2.7$\times$10$^{-30}$ &  &  \\
& & &  & & (\ce{CO2}) 2.9$\times$10$^{-30}$ & & \\
& & &  & & (\ce{H2}) 5.1$\times$10$^{-30}$ & & \\
& & &  & & & & \\
12. & \ce{OH + OH + M -> H2O2 + M} & 2.3$\times$10$^{-11}$ & 1.5--6.5$\times$10$^{-11}$ & 1 & (\ce{N2}) 4.9$\times$10$^{-32}$ & (\ce{N2}) 5.1--330$\times$10$^{-32}$ & 1 \\
& & &  & & (\ce{CO2}) 5.5$\times$10$^{-32}$ & (\ce{CO2}) 6.4--420$\times$10$^{-32}$ & 1 \\
& & &  & & (\ce{H2}) 9.9$\times$10$^{-32}$ & \footnotesize (He,\ce{H2O}) 1.3--1800$\times$10$^{-32}$ & 1 \\
& & &  & & & & \\
*13. & \ce{OH + ^3O + M -> HO2 + M} & 7.4$\times$10$^{-11}$ & & & (\ce{N2}) 8.5$\times$10$^{-32}$ &  &  \\
& & &  & & (\ce{CO2}) 9.4$\times$10$^{-32}$ & & \\
& & &  & & (\ce{H2}) 1.8$\times$10$^{-31}$ & & \\
& & &  & & & & \\
*14. & \ce{OH + ^1O + M -> HO2 + M} & 1.0$\times$10$^{-9}$ & & & (\ce{N2}) 4.1$\times$10$^{-30}$ &  &  \\
& & &  & & (\ce{CO2}) 4.5$\times$10$^{-30}$ & & \\
& & &  & & (\ce{H2}) 8.3$\times$10$^{-30}$ & & \\
& & &  & & & & \\
15. & \ce{OH + NH + M -> } & 7.0$\times$10$^{-12}$ & & & (\ce{N2}) 8.5$\times$10$^{-31}$ & & \\
& \ce{OH$\cdots$NH* + M* ->} & & & & (\ce{CO2}) 9.2$\times$10$^{-31}$ & & \\
& \ce{trans-HNOH + M} & &  & & (\ce{H2}) 1.7$\times$10$^{-30}$ & & \\
& & &  & & & & \\
16. & \ce{OH + CH3 + M -> } & 2.0$\times$10$^{-11}$ & 9.3--17$\times$10$^{-11}$ & 5 & (\ce{N2}) 2.1$\times$10$^{-27}$ & \footnotesize (He,\ce{SF6}) 2.0--7.2$\times$10$^{-27}$ & 1 \\
& \ce{OH$\cdots$CH3* + M* -> } & &  & & (\ce{CO2}) 2.3$\times$10$^{-27}$ & " & 1 \\
& \ce{CH3OH + M} & &  & & (\ce{H2}) 3.8$\times$10$^{-27}$ & " & 1 \\
& & &  & & & & \\
17. & \ce{OH + H + M -> H2O + M} & 2.4$\times$10$^{-10}$ & & & (\ce{N2}) 3.0$\times$10$^{-31}$ & (\ce{N2}) 4.8--6.8$\times$10$^{-31}$ & 2 \\
& & &  & & (\ce{CO2}) 3.7$\times$10$^{-31}$ & (\ce{CO2}) 9.0$\times$10$^{-31}$ & 2 \\
& & &  & & (\ce{H2}) 5.1$\times$10$^{-31}$ & \footnotesize (He,\ce{H2O}) 1.5--6.8$\times$10$^{-31}$ & 1 \\
& & &  & & & & \\
*18. & \ce{^3O + CN + M -> NCO + M} & 7.1$\times$10$^{-12}$ & 9.4--16$\times$10$^{-12}$ & 1 & (\ce{N2}) 1.3$\times$10$^{-30}$ & & \\
& & &  & & (\ce{CO2}) 1.5$\times$10$^{-30}$ & & \\
& & &  & & (\ce{H2}) 2.6$\times$10$^{-30}$ & & \\
& & &  & & & & \\
19. & \ce{^3O + ^3O + M -> O2 + M} & 1.8$\times$10$^{-11}$ & & & (\ce{N2}) 3.0$\times$10$^{-34}$ & (\ce{N2}) 3.1--10$\times$10$^{-33}$ & 10 \\
& & &  & & (\ce{CO2}) 3.2$\times$10$^{-34}$ & \footnotesize  (\ce{Ar,O2}) 3.9--100$\times$10$^{-34}$ & 1 \\
& & &  & & (\ce{H2}) 6.1$\times$10$^{-34}$ & \footnotesize (\ce{Ar,N2}) 3.9--100$\times$10$^{-34}$ & 1 \\
& & &  & & & & \\
20. & \ce{^3O + ^4N + M -> NO + M} & 6.6$\times$10$^{-11}$ & & & (\ce{N2}) 1.6$\times$10$^{-33}$ & (\ce{N2}) 5--11$\times$10$^{-33}$ & 3 \\
& & &  & & (\ce{CO2}) 1.8$\times$10$^{-33}$ & (\ce{CO2}) 1.8$\times$10$^{-32}$ & 10\\
& & &  & & (\ce{H2}) 3.3$\times$10$^{-33}$ &\footnotesize (\ce{He,N2}) 3.8--11$\times$10$^{-33}$ & 1\\
& & &  & & & & \\
*21. & \ce{^3O + ^3CH2 + M -> H2CO + M} & 6.7$\times$10$^{-11}$ & 1.9--20$\times$10$^{-11}$ & 1 & (\ce{N2}) 9.2$\times$10$^{-29}$ & & \\
& & &  & & (\ce{CO2}) 1.1$\times$10$^{-28}$ & & \\
& & &  & & (\ce{H2}) 1.7$\times$10$^{-28}$ & & \\
& & &  & & & & \\
*22. & \ce{^3O + CH + M -> HCO + M} & 1.1$\times$10$^{-10}$ & 6.6--9.5$\times$10$^{-11}$ & 1 & (\ce{N2}) 5.2$\times$10$^{-30}$ & & \\
& & &  & & (\ce{CO2}) 6.2$\times$10$^{-30}$ & & \\
& & &  & & (\ce{H2}) 9.9$\times$10$^{-30}$ & & \\
& & &  & & & & \\
23. & \ce{^3O + H + M -> OH + M} & 3.5$\times$10$^{-10}$ & & & (\ce{N2}) 2.6$\times$10$^{-33}$ & (\ce{M}) 1--8000$\times$10$^{-33}$ & 1 \\
& & &  & & (\ce{CO2}) 2.9$\times$10$^{-33}$ & " & 1 \\
& & &  & & (\ce{H2}) 4.6$\times$10$^{-33}$ & " & 1 \\
& & &  & & & & \\
*24. & \ce{^1O + HCN + M -> HCNO + M} & 3.3$\times$10$^{-11}$ & & & (\ce{N2}) 4.0$\times$10$^{-29}$ & & \\
& & &  & & (\ce{CO2}) 4.6$\times$10$^{-29}$ & & \\
& & &  & & (\ce{H2}) 8.0$\times$10$^{-29}$ & & \\
& & &  & & & & \\
*25. & \ce{^1O + CN + M -> NCO + M} & 8.9$\times$10$^{-11}$ & & & (\ce{N2}) 1.9$\times$10$^{-29}$ & & \\
& & &  & & (\ce{CO2}) 2.1$\times$10$^{-29}$ & & \\
& & &  & & (\ce{H2}) 3.6$\times$10$^{-29}$ & & \\
& & &  & & & & \\
*26. & \ce{^1O + ^1O + M -> O2 + M} & 2.3$\times$10$^{-10}$ & & & (\ce{N2}) 8.8$\times$10$^{-33}$ & & \\
& & &  & & (\ce{CO2}) 9.6$\times$10$^{-33}$ & & \\
& & &  & & (\ce{H2}) 1.8$\times$10$^{-32}$ & & \\
& & &  & & & & \\
*27. & \ce{^1O + CH4 + M -> CH3OH + M} & 5.8$\times$10$^{-9}$ & 1.4--4.0$\times$10$^{-10}$ & 15 & (\ce{N2}) 3.6$\times$10$^{-23}$ & & \\
& & &  & & (\ce{CO2}) 3.9$\times$10$^{-23}$ & & \\
& & &  & & (\ce{H2}) 6.3$\times$10$^{-23}$ & & \\
& & &  & & & & \\
*28. & \ce{^1O + ^1CH2 + M -> H2CO + M} & 3.3$\times$10$^{-10}$ & & & (\ce{N2}) 6.6$\times$10$^{-27}$ & & \\
& & &  & & (\ce{CO2}) 7.7$\times$10$^{-27}$ & & \\
& & &  & & (\ce{H2}) 1.2$\times$10$^{-26}$ & & \\
& & &  & & & & \\
*29. & \ce{^1O + CH + M -> HCO + M} & 9.2$\times$10$^{-11}$ & & & (\ce{N2}) 4.9$\times$10$^{-29}$ & & \\
& & &  & & (\ce{CO2}) 5.8$\times$10$^{-29}$ & & \\
& & &  & & (\ce{H2}) 9.1$\times$10$^{-29}$ & & \\
& & &  & & & & \\
*30. & \ce{^1O + H2 + M -> H2O + M} & 7.1$\times$10$^{-10}$ & 1.1--3.0$\times$10$^{-10}$ & 2 & (\ce{N2}) 1.2$\times$10$^{-29}$ & & \\
& & &  & & (\ce{CO2}) 1.4$\times$10$^{-29}$ & & \\
& & &  & & (\ce{H2}) 2.0$\times$10$^{-29}$ & & \\
& & &  & & & & \\
*31. & \ce{^1O + H + M -> OH + M} & 1.1$\times$10$^{-9}$ & & & (\ce{N2}) 1.4$\times$10$^{-32}$ & & \\
& & &  & & (\ce{CO2}) 1.5$\times$10$^{-32}$ & & \\
& & &  & & (\ce{H2}) 2.3$\times$10$^{-32}$ & & \\
\hline
\multicolumn{8}{l}{\footnotesize $*$ Reactions with no previously known rate coefficients.} \\
\end{longtable*}

Our calculated high-pressure rate coefficients are within the range of experimental values in 9 out of 12 cases. The other three rate coefficients are factors of 2, 5, and 15 from the nearest experimental values. Typical uncertainties for rate coefficients--as assigned in large experimental data evaluations--range from factors of 2--10 \citep{Reference451,Reference509}; Therefore, this calculated accuracy is consistent with the levels of uncertainty typically found in the literature.

Each low-pressure limit rate coefficient was calculated for three bath gases (\ce{N2}, \ce{CO2}, and \ce{H2}) and compared to experiments performed with matching bath gases when possible, and any bath gases otherwise. Nine of the reactions had experimentally measured low-pressure limit rate coefficients for one or more bath gases. All of our calculated rate coefficients for these reactions landed within an order of magnitude of the experimental range for the matching bath gas when possible, or another bath gas otherwise. Most commonly (67\% of the time), our rate coefficients were within a factor of 3 from the nearest experimental measurement. Larger deviations tended to occur for cases that only have a single experimental measurement for comparison.

In Table~\ref{Results2}, we display the 95 one- and two-body reaction rate coefficients calculated at 298K with any experimental or suggested values. 47 of these reactions have experimental or suggested values, and our calculations are within approximately one order of magnitude of these values in all but one case. In 60\% of cases our calculated rate coefficients are within a factor of 2 of experimental values, and in 83\% of cases our calculated rate coefficients are within a factor of 6 of experimental values. 

In one case, \ce{OH + CH4 -> H2O + CH3}, our calculated rate coefficient has a slightly higher than an order of magnitude deviation from experiment (factor of 54). We attribute this error to the lack of a quantum tunneling correction in our calculations. \citet{Bravo-Perez2005} performed transition state theory calculations for this reaction at the BHandHLYP/6-311G(d,p) level of theory, and calculated a tunneling factor of 30.56 at 298 K using an Eckart model. If we applied this factor to our calculation, our rate coefficient would be within a factor of two of the experimental range.

\setlength\LTcapwidth{\textwidth}
\begin{longtable*}{clccccc}
\caption{Calculated reaction rate coefficients at 298 K for the one- and two-body reactions in this paper. Calculations are performed at the BHandHLYP/aug-cc-pVDZ level of theory. Reactions with rate coefficients slower than k = 10$^{-21}$ are not included in this network. The precense or absence of an energy barrier in the rate-limiting step (or the only step) of the reaction is specified. The error factor is the multiplicative or divisional factor from the nearest experimental or suggested value; the error factor is 1 if the calculated value is within the range of experimental or suggested values. First-order rate coefficients have units s$^{-1}$. Second-order rate coefficients have units cm$^{3}$s$^{-1}$. \label{Results2}} \\
No. & Reaction equation & Forw./Rev. & Barrier? & k(298) calculated & k(298) experimental & Error factor\\ \hline \\[-2mm]
*32. & \ce{NCCO -> CO + CN} & F & Y & 9.4$\times$10$^{-12}$ &  & \\
33. & \ce{CO2 + ^1O -> ^1CO3* -> ^3CO3* ->} & F & N & 3.8$\times$10$^{-11}$ & 0.1--23$\times$10$^{-11}$ & 1 \\ 
 & \ce{CO2 + ^3O} & & & & & \\
34. & \ce{CO2 + ^2N -> NCO2* -> OCNO* ->} & F & $^a$Y & 3.2$\times$10$^{-14}$ & 1.8--6.8$\times$10$^{-13}$ & 6 \\
 & \ce{CO + NO} & & & & & \\
35. & \ce{CO2 + ^1CH2 -> ^1CH2CO2* ->} & F & N & 8.0$\times$10$^{-13}$ &  &  \\
 & \ce{H2CO + CO} & & & & & \\
36. & \ce{CO2 + CH -> CHCO2* -> HCOCO* ->} & F & $^b$N & 3.1$\times$10$^{-12}$ & 1.8--2.1$\times$10$^{-12}$ &  1 \\
 & \ce{HCO + CO} & & & & & \\
37. & \ce{H2O2 -> OH + OH} & F & Y & 5.1$\times$10$^{-9}$ & & \\
38. & \ce{H2CO + CN -> HCN + HCO} & F & N & 1.7$\times$10$^{-11}$ & 1.7$\times$10$^{-11}$ & 1 \\
39. & \ce{H2CO + OH -> r,l-H2COHO* ->} & F & Y & 7.1$\times$10$^{-17}$ &  &  \\
 & \ce{trans-HCOHO* + H* -> H2O + CO + H} & & & & & \\
40. & \ce{H2CO + OH -> H2CO$\cdots$HO* -> H2O + HCO} & F & Y & 1.1$\times$10$^{-12}$ & 6.1--15$\times$10$^{-12}$ & 6 \\
41. & \ce{H2CO + ^3O -> HCO + OH} & F & Y & 6.8$\times$10$^{-14}$ & 1.5--1.9$\times$10$^{-13}$ & 2 \\
*42. & \ce{H2CO + ^1O -> H2CO2* -> HCO2H* ->} & F & N & 4.6$\times$10$^{-10}$ &  &  \\
 & \ce{HCO + OH} & & & & & \\
43. & \ce{H2CO + CH3 -> HCO + CH4} & F & Y & 1.9$\times$10$^{-19}$ &  2.2--4.2$\times$10$^{-18}$ & 12 \\
44. & \ce{H2CO + ^3CH2 -> HCO + CH3} & F & Y & 1.1$\times$10$^{-14}$ & $<$1.0$\times$10$^{-14}$ & 1 \\
45. & \ce{H2CO + ^1CH2 -> HCO + CH3} & F & N & 1.5$\times$10$^{-12}$ & 2.0$\times$10$^{-12}$ & 1 \\
46. & \ce{H2CO + CH -> H2COCH_a* ->} & F & N & 3.1$\times$10$^{-11}$ & 3.8$\times$10$^{-10}$ & 12 \\
 & \ce{H2COCH_b* -> CH2HCO* -> } & & & & & \\
 & \ce{CH3CO* -> CO + CH3} & & & & & \\
47. & \ce{H2CO + CH -> H2COCH_c* ->} & F & N & 1.1$\times$10$^{-12}$ &  & \\
& \ce{HCO + ^3CH2} & & & & & \\
48. & \ce{H2CO + H -> HCO + H2} & F & Y & 1.8$\times$10$^{-13}$ & 3.9--6.7$\times$10$^{-14}$ & 3 \\
*49. & \ce{HCO + H2CN -> H2CO + HCN} & F & Y & 7.0$\times$10$^{-15}$ &  &  \\
50. & \ce{HCO + HCO -> trans-C2H2O2* ->} & F & Y & 1.2$\times$10$^{-13}$ & 2.8--750$\times$10$^{-13}$ & 2 \\ 
 & \ce{anti-HCOH* + CO* -> H2CO + CO} & & & & & \\
51. & \ce{HCO + HCO -> cis-C2H2O2* ->} & F & N & 7.4$\times$10$^{-11}$ & 3.6$\times$10$^{-11}$ & 2 \\
 & \ce{CO + CO + H2} & & & & & \\
52. & \ce{HCO + CN -> HCOCN* -> CO + HCN} & F & N & 5.4$\times$10$^{-12}$ &  & \\
53. & \ce{HCO + OH -> trans-HCOHO* -> CO + H2O} & F & N & 7.0$\times$10$^{-12}$ & 5--18$\times$10$^{-11}$ & 7 \\
54. & \ce{HCO + ^3O -> HCO2* -> CO2 + H} & F & N & 2.6$\times$10$^{-11}$ & 5.0$\times$10$^{-11}$ & 2\\
55. & \ce{HCO + ^3O -> CO + OH} & F & N & 3.4$\times$10$^{-11}$ & 5.0$\times$10$^{-11}$ & 1\\
*56. & \ce{HCO + ^1O -> HCO2* -> CO2 + H} & F & N & 1.5$\times$10$^{-10}$ &  & \\
*57. & \ce{HCO + NH -> H2CO + ^4N} & F & Y & 3.6$\times$10$^{-20}$ &  &  \\
*58. & \ce{HCO + NH -> CO + NH2} and & F & N & 1.4$\times$10$^{-11}$ &  & \\
 & \ce{HCO + NH -> HNHCO* -> H2NCO* ->} & & & & & \\
& \ce{CO + NH2} & & & & & \\
59. & \ce{HCO + ^4N -> ^3NCOH* -> NCO + H} & F & N & 2.8$\times$10$^{-11}$ &  & \\
60. & \ce{HCO + ^4N -> CO + NH} & F & N & 2.2$\times$10$^{-11}$ &  & \\
*61. & \ce{HCO + ^2N -> ^3NCOH* -> NCO + H} & F & N & 6.6$\times$10$^{-11}$ &  & \\
*62. & \ce{HCO + ^2N -> CO + NH} & F & N & 4.8$\times$10$^{-11}$ &  & \\
63. & \ce{HCO + CH3 -> CO + CH4} & F & N & 1.0$\times$10$^{-11}$ & 3.6$\times$10$^{-11}$--2.0$\times$10$^{-10}$ & 4\\
64. & \ce{HCO + ^3CH2 -> CH3 + CO} and & F & N & 2.1$\times$10$^{-11}$ & 3.0$\times$10$^{-11}$ & 1\\
 & \ce{HCO + ^3CH2 -> CH2HCO* ->} & & & & & \\
 &  \ce{CH3CO* -> CH3 + CO} & & & & & \\
65. & \ce{HCO + ^1CH2 -> CH2HCO* ->} & F & N & 1.2$\times$10$^{-11}$ & 3.0$\times$10$^{-11}$ & 3 \\
 & \ce{CH3CO* -> CH3 + CO} & & & & & \\
66. & \ce{HCO + CH -> CO + ^3CH2} & F & N & 1.5$\times$10$^{-11}$ &  & \\
*67. & \ce{HCO + CH -> CO + ^1CH2} & F & N & 4.6$\times$10$^{-12}$ &  & \\
68. & \ce{HCO + H -> CO + H2} and & F & N & 6.9$\times$10$^{-11}$ & 1.1--5.5$\times$10$^{-10}$ & 2\\
 & \ce{HCO + H -> H2CO$_{(\nu)}$* -> CO + H2} & & & & & \\
69. & \ce{HCO + H -> H2CO$_{(\nu)}$* -> CO + H + H} & F & N & 2.4$\times$10$^{-11}$ &  & \\
70. & \ce{HCO -> CO + H} & F & Y & 2.2$\times$10$^{-2}$ &  &  \\
71. & \ce{CO + OH -> OH$\cdots$CO* -> cis-HOCO* -> } & F & Y & $^c$2.9$\times$10$^{-12}$ & 0.9--9.7$\times$10$^{-13}$ & 3 \\
 & \ce{CO2 + H} & & & & & \\
72. & \ce{H2O + ^1O -> H2OO* -> H2O2* ->} & F & N & 4.8$\times$10$^{-10}$ & 1.8--3.7$\times$10$^{-10}$ & 1 \\
 & \ce{OH + OH} & & & & & \\
73. & \ce{H2O + CN -> H2OCN* -> OH + HCN} & F & Y & 6.6$\times$10$^{-15}$ &  &  \\
74. & \ce{H2O + ^2N -> H2ON* -> trans-HNOH* ->} & F & N & 1.9$\times$10$^{-10}$ & & \\
 & \ce{HNO + H} and & & & & & \\
 & \ce{H2O + ^2N -> H2ON* -> trans-HNOH* ->} & & & & & \\
 & \ce{H2NO* -> HNO + H} & & & & & \\
75. & \ce{H2O + CH -> H2O$\cdots$CH* -> H2OCH* ->} & F & $^d$N & 2.0$\times$10$^{-10}$ & 1.3--4.5$\times$10$^{-11}$ & 4 \\
 & \ce{H2COH* -> H2CO + H} & & & & & \\
*76. & \ce{H2O + CH -> OH + ^3CH2} & F & Y & 3.9$\times$10$^{-16}$ & & \\
77. & \ce{OH + HCN -> NCHOH* -> HOCN + H} & F & Y & 1.2$\times$10$^{-15}$ & 0.1--31$\times$10$^{-15}$ & 1 \\
78. & \ce{OH + CN -> HO$\cdots$CN -> ^3HOCN_1* -> } & F & Y & 1.1$\times$10$^{-12}$ &  &  \\
 & \ce{^3HOCN_2 -> NCO + H} & & & & & \\
79. & \ce{OH + CN -> HCN + ^3O} & F & Y & 4.5$\times$10$^{-13}$ &  &  \\
*80. & \ce{OH + CN -> HNC + ^3O} & F & Y & 2.3$\times$10$^{-17}$ &  &  \\
81. & \ce{OH + OH -> trans-^3H2O2* -> H2O + ^3O} & F & N & $^e$2.5$\times$10$^{-11}$ & 0.8--2.6$\times$10$^{-12}$ & 10 \\
82. & \ce{OH + ^3O -> HO2$_{(\nu)}$* -> O2 + H} & F & N & 7.4$\times$10$^{-11}$ & 2.8--4.2$\times$10$^{-11}$ & 2 \\
*83. & \ce{OH + ^1O -> HO2$_{(\nu)}$* -> O2 + H} & F & N & 1.0$\times$10$^{-9}$ &  &  \\
84. & \ce{OH + NH -> OH$\cdots$NH* -> trans-HNOH* ->} & F & $^f$N & 7.0$\times$10$^{-12}$ & 3.3$\times$10$^{-11}$ & 5 \\
 & \ce{HNO + H} and & & & & & \\
 & \ce{OH + NH -> OH$\cdots$NH* -> trans-HNOH* ->} & & & & & \\ 
 & \ce{H2NO* -> HNO + H} & & & & & \\
85. & \ce{OH + NH -> H2O + ^4N} & F & Y & 6.8$\times$10$^{-13}$ & 3.1$\times$10$^{-12}$ & 5 \\
86. & \ce{OH + ^4N -> ^3OH$\cdots$N -> ^3NOH* ->} & F & Y & 1.0$\times$10$^{-10}$ & 4.2--5.3$\times$10$^{-11}$ & 2 \\
 & \ce{NO + H} & & & & & \\
*87. & \ce{OH + ^2N -> ^3OH$\cdots$N -> ^3NOH* ->} & F & N & 1.5$\times$10$^{-10}$ &  &  \\
 & \ce{NO + H} & & & & \\
88. & \ce{OH + CH4 -> H2O + CH3} & F & Y & 1.1$\times$10$^{-16}$ & 5.9--11$\times$10$^{-15}$ & 54 \\
89. & \ce{OH + CH3 -> ^3O + CH4} & F & Y & 1.1$\times$10$^{-18}$ & 1.8$\times$10$^{-17}$ & 16 \\
90. & \ce{OH + CH3 -> H2O + ^3CH2} & F & Y & 3.5$\times$10$^{-18}$ &  &  \\
91. & \ce{OH + ^3CH2 -> OH$\cdots$CH2* -> H2COH* -> } & F & N & 4.6$\times$10$^{-11}$ & 3.0$\times$10$^{-11}$ & 2 \\
& \ce{H2CO + H} & & & & & \\
92. & \ce{OH + ^3CH2 -> H2O + CH} & F & N & 7.6$\times$10$^{-13}$ &  &  \\
93. & \ce{OH + ^1CH2 -> OH$\cdots$CH2* -> H2COH* ->} & F & N &  4.6$\times$10$^{-11}$ & 5.0$\times$10$^{-11}$ & 1 \\
& \ce{H2CO + H} & & & & & \\
94. & \ce{OH + CH -> ^3OH$\cdots$CH* -> ^3HCOH* -> } & F & N & 3.2$\times$10$^{-11}$ &  &  \\ 
& \ce{^3H2CO* -> HCO + H} & &  &  &  & \\
95. & \ce{OH + CH -> anti-HCOH$_{(\nu)}$* ->} & F & N & $^g$6.3$\times$10$^{-12}$ &  &  \\ 
& \ce{H2CO$_{(\nu)}$* -> CO + H2} & &  &  &  & \\
96. & \ce{OH + CH -> anti-HCOH$_{(\nu)}$* ->} &  F & N & $^g$6.3$\times$10$^{-12}$ &  &  \\
& \ce{H2CO$_{(\nu)}$* -> CO + H + H} & &  &  &  & \\
97. & \ce{OH + H2 -> H2O + H} & F & Y & 1.5$\times$10$^{-15}$ & 5.3--8.5$\times$10$^{-15}$ & 4 \\
98. & \ce{OH + H -> ^3O + H2} & F & Y & 6.5$\times$10$^{-16}$ & 9.9$\times$10$^{-17}$--5.6$\times$10$^{-16}$ & 1 \\
*99. & \ce{^3O + H2CN -> CH2NO* ->} & F & Y & 4.0$\times$10$^{-14}$ &  &  \\
 & \ce{HCNO + H} & &  &  &  & \\
*100. & \ce{^3O + H2CN <- CH2NO* <-} & R & N & 9.8$\times$10$^{-11}$ & & \\ 
 & \ce{HCNO + H} & &  &  &  & \\
101. & \ce{^3O + H2CN -> CH2NO* ->} & F & Y & 8.3$\times$10$^{-15}$ &  &  \\ 
 & \ce{HCNOH* -> OH + HCN} & & & & & \\
102. & \ce{^3O + HCN -> ^3NCOH -> NCO + H} & F & Y & 1.5$\times$10$^{-18}$ &  &  \\
103. & \ce{^3O + HCN <- ^3NCOH <- NCO + H} & R & Y & 2.5$\times$10$^{-20}$ &  & \\
104. & \ce{^3O + CN -> ^4NCO -> CO + ^4N} & F & N & 1.5$\times$10$^{-11}$ & 2.7$\times$10$^{-12}$--3.7$\times$10$^{-11}$ & 1 \\
105. & \ce{^3O + CN -> NCO$_{(\nu)}$ -> CO + ^2N} & F & N & 7.1$\times$10$^{-12}$ & 9.4$\times$10$^{-12}$--1.6$\times$10$^{-11}$ & 1 \\
106. & \ce{^3O + NH -> HNO* -> NO + H} & F & N & 3.1$\times$10$^{-11}$ & 5.0$\times$10$^{-11}$ & 2 \\
107. & \ce{^3O + NH -> OH + ^4N} & F & Y & 2.2$\times$10$^{-14}$ & $<$1.7$\times$10$^{-13}$--5.0$\times$10$^{-12}$ & 1	 \\
108. & \ce{^3O + CH4 -> OH + CH3} & F & Y & 1.1$\times$10$^{-19}$ & 6.6$\times$10$^{-19}$--6.6$\times$10$^{-16}$ & 6 \\
109. & \ce{^3O + CH3 -> CH3O* -> H2CO + H} & F & N & 9.4$\times$10$^{-11}$ & $>$3.0$\times$10$^{-11}$--1.9$\times$10$^{-10}$ & 1 \\
110. & \ce{^3O + ^3CH2 -> H2CO$_{(\nu)}$ -> CO + H + H} & F & N & 3.4$\times$10$^{-11}$ & $^h$1.0$\times$10$^{-11}$--1.0$\times$10$^{-10}$ & 1 \\
111. & \ce{^3O + ^3CH2 -> H2CO$_{(\nu)}$ -> CO + H2} & F & N & 3.4$\times$10$^{-11}$ & $^h$1.0$\times$10$^{-11}$--1.0$\times$10$^{-10}$ & 1 \\
*112. & \ce{^3O + ^1CH2 -> ^3H2CO* -> HCO + H} & F & N & 2.1$\times$10$^{-10}$ &  &  \\
113. & \ce{^3O + CH -> HCO$_{(\nu)}$ -> CO + H} & F & N & 1.1$\times$10$^{-10}$ & 6.6$\times$10$^{-11}$ & 2 \\
114. & \ce{^3O + CH -> ^4HCO* -> ^4COH* ->} & F & N & 2.5$\times$10$^{-10}$ &  &  \\
& \ce{OH + C} & & & & & \\
115. & \ce{^3O + H2 -> OH + H} & F & Y & 7.2$\times$10$^{-19}$ & 7.0$\times$10$^{-18}$--1.1$\times$10$^{-17}$ & 10 \\
*116. & \ce{^1O + H2CN -> CH2NO* ->} & F & Y & 4.5$\times$10$^{-10}$ & & \\ 
 & \ce{^3O + H2CN} & &  &  &  & \\
*117. & \ce{^1O + H2CN -> CH2NO* ->} & F & Y & 6.0$\times$10$^{-13}$ & & \\ 
 & \ce{HCNO + H} & &  &  &  & \\
*118. & \ce{^1O + H2CN -> CH2NO* ->} & F & Y & 1.2$\times$10$^{-13}$ & & \\ 
 & \ce{HCNOH* -> HCN + OH} & &  &  &  & \\
*119. & \ce{^1O + CN -> NCO$_{(\nu)}$ -> CO + ^2N} & F & N & 8.9$\times$10$^{-11}$ & &  \\
120. & \ce{^1O + CH4 -> CH3OH$_{(\nu)}$ -> OH + CH3} & F & N & 5.8$\times$10$^{-9}$ & 1.4--4.0$\times$10$^{-10}$ & 15 \\
*121. & \ce{^1O + CH3 -> CH3O* -> H2CO + H} & F & N & 4.3$\times$10$^{-10}$ &  & \\
*122. & \ce{^1O + ^3CH2 -> ^3H2CO* -> HCO + H} & F & N & 7.0$\times$10$^{-10}$ &  & \\
*123. & \ce{^1O + ^1CH2 -> H2CO$_{(\nu)}$ ->} & F & N & 1.7$\times$10$^{-10}$ &  & \\
 & \ce{CO + H + H} & & & & & \\
*124. & \ce{^1O + ^1CH2 -> H2CO$_{(\nu)}$ -> CO + H2} & F & N & 1.7$\times$10$^{-10}$ &  & \\
*125. & \ce{^1O + CH -> HCO$_{(\nu)}$ -> CO + H} & F & N & 9.2$\times$10$^{-11}$ &  &  \\
126. & \ce{^1O + H2 -> H2O$_{(\nu)}$ -> OH + H} & F & N & 7.1$\times$10$^{-10}$ & 1.1--3.0$\times$10$^{-10}$ & 2\\
\hline
\multicolumn{7}{l}{\footnotesize $^a$ We introduce a barrier of 17.15 kJ mol$^{-1}$ (half the HF barrier) to this calculation as no barrier is found at the BHandHLYP/aug-cc-pVDZ level} \\
\multicolumn{7}{l}{\footnotesize of theory (see supplement for more details).} \\
\multicolumn{7}{l}{\footnotesize $^b$ We remove the barrier from this calculation as experiment predicts this reaction to be barrierless below 400 K \citep{Reference1018}.} \\
\multicolumn{7}{l}{\footnotesize $^c$ We remove the intermediate barriers from this reaction and reduce the barrierless first step by a factor of 3.4 to match the barrier effects at the } \\
\multicolumn{7}{l}{\footnotesize B3LYP/aug-cc-pVDZ level of theory. Experiments predict this reaction to have little to no barrier \citep{Reference451}. } \\
\multicolumn{7}{l}{\footnotesize $^d$ We remove the barrier from the rate limiting third step of this calculation, as experiment predicts this reaction to be barrierless \citep{Reference1193}. } \\
\multicolumn{7}{l}{\footnotesize $^e$ Simulations did not converge beyond a O-O bond distance of 2.90$\AA$. The rate coefficient is calculated with the variational transition state at} \\
\multicolumn{7}{l}{\footnotesize this location, which has the highest $\Delta$G.} \\
\multicolumn{7}{l}{\footnotesize $^f$ We remove the barrier from the rate limiting third step of this calculation, as data evaluations suggest little to no barrier for this reaction \citep{Reference1053}. } \\
\multicolumn{7}{l}{\footnotesize $^g$ This rate coefficient is one half of the calculated rate coefficient for \ce{OH + CH -> anti-HCOH$_{(\nu)}$*} as both \ce{CO + H2} and \ce{CO + H + H} are} \\
\multicolumn{7}{l}{\footnotesize equally probable decay pathways for \ce{anti-HCOH$_{(\nu)}$} \citep{Reference1074,Reference451,Reference509}.} \\
\multicolumn{7}{l}{\footnotesize $^h$ Experimental values are for \ce{^3O + ^3CH2 -> products} divided by 2. As both product channels \ce{CO + H + H} and \ce{CO + H2} are suggested to be} \\
\multicolumn{7}{l}{\footnotesize equally likely \citep{Reference1074,Reference451,Reference509}.} \\
\end{longtable*}

\subsubsection*{Method Limitations}

Occasionally computational methods misdiagnose reaction energy barriers. In other words, a method may calculate a barrier when experiments suggest the reaction is barrierless, or a method may calculate no barrier when experiments suggest a small-to-modest-sized barrier ($\sim$1--20 kJ mol$^{-1}$) exists. We find this to be biggest limitation of applying a consistent computational quantum method to a large number of reactions. This is the main reason for taking a hybrid approach to building CRAHCN-O. Experiments are the most accurate method to calculate rate coefficients, therefore experimental values will always be used when possible. However, for the large number of reactions without experimentally measured rate coefficients, we must use a robust and feasible computational method to calculate and include these reactions in the network. 

In four cases (noted in Table~\ref{Results2}), our chosen computational method (BHandHLYP/aug-cc-pVDZ) predicts barriers at the first step or an intermediate step of reactions that are expected to be barrierless. In one other case, this method predicts a reaction had no barrier, when experiment suggests a barrier of 17.15 kJ mol$^{-1}$ \citep{Reference557}. For these few cases, we artificially remove the barriers from these calculations, or introduce an experimental barrier. Based on the calculations in this paper, we find this method correctly diagnoses barriers $\sim$92\% of the time.

Comparing the barrier diagnosis capabilities of BHandHLYP/aug-cc-pVDZ with two other widely used method in past work \citep{Pearce2020a}, we find CCSD/aug-cc-pVTZ and $\omega$B97XD/aug-cc-pVDZ share these limitations. For 11 chosen reactions, BHandHLYP/aug-cc-pVDZ misdiagnosed 4 barriers, CCSD/aug-cc-pVTZ misdiagnosed 5 barriers, and $\omega$B97XD/aug-cc-pVDZ misdiagnosed 2 barriers.

A second limitation of our method is that we do not include a correction factor for quantum mechanical tunneling. This may not be a big concern at 298 K, where our rate coefficient calculations are typically within a factor of two of experimental values, and generally always within an order of magnitude of experimental values. However, tunneling is most relevant at lower temperatures \citep{Meisner2016}. 

Given the lack of experimental low temperature ($\lesssim$ 230 K) rate coefficient data for the reactions in this study, we cannot obtain a valid statistical sense of the accuracy of our method for calculating low temperature rate coefficients. However, it is a reasonable assumption that our treatment leads to larger uncertainties at the lower end of our temperature range (50--200 K), where tunneling plays a greater role; possibly up to two orders of magnitude.

\section*{Discussion}\label{discussion}

\subsection*{Highlighted New Reactions}

As we have already noted, we have discovered 45 previously unknown reactions and provide the first calculations of their rate coefficients. In Table~\ref{Highlights}, we highlight 6 of these reactions. These reactions are potentially key pathways for the production and destruction of HCN or \ce{H2CO} in planetary atmospheres due to their high rate coefficients at 298 K, and the reasonably high abundances of their reactants in atmospheres.

\begin{table*}[t]
\centering
\caption{Highlighted newly discovered reactions in this work, listed with their calculated rate coefficients at 298 K and potential for importance in atmospheres. For simplicity, reaction intermediates are not listed here. See Tables \ref{Results1} and \ref{Results2} for full details of reaction
intermediates. Second-order rate coefficients have units cm$^{3}$s$^{-1}$. Third-order rate coefficients have units cm$^{6}$s$^{-1}$. \label{Highlights}} 
\begin{tabular}{lrl}
\\
\multicolumn{1}{c}{Reaction} & 
\multicolumn{1}{c}{\hspace{0.3cm}k(298) calculated} & 
\multicolumn{1}{c}{\hspace{0.3cm}Importance}\\ \hline \\[-2mm]
\ce{^1O + CH3 -> H2CO + H} & 4.3$\times$10$^{-10}$ & \ce{H2CO} production in upper atmospheres \\
 & & \\
\ce{^1O + ^1CH2 + M -> H2CO + M} & k$_{\infty}$ = 3.3$\times$10$^{-10}$  & \ce{H2CO} production in lower atmospheres \\
 & k$_0$(\ce{N2}) = 6.6$\times$10$^{-27}$ & \\
  & k$_0$(\ce{CO2}) = 7.7$\times$10$^{-27}$ & \\
   & k$_0$(\ce{H2}) = 1.2$\times$10$^{-26}$ & \\
 & & \\
   \ce{^3O + ^3CH2 + M -> H2CO + M} & k$_{\infty}$ = 6.7$\times$10$^{-11}$  & \ce{H2CO} production in lower atmospheres \\
 & k$_0$(\ce{N2}) = 9.2$\times$10$^{-29}$ & \\
  & k$_0$(\ce{CO2}) = 1.1$\times$10$^{-28}$ & \\
   & k$_0$(\ce{H2}) = 1.7$\times$10$^{-28}$ & \\
 & & \\
   \ce{^1O + H2CN -> HCN + OH} & 1.2$\times$10$^{-13}$  & \ce{HCN} production in upper atmospheres \\
 & & \\
   \ce{H2CO + ^1O -> HCO + OH} & 4.6$\times$10$^{-10}$ & \ce{H2CO} destruction in upper atmospheres \\
 & & \\
   \ce{^1O + HCN + M -> HCNO + M} & k$_{\infty}$ = 3.3$\times$10$^{-11}$  & \ce{HCN} destruction in lower atmospheres \\
 & k$_0$(\ce{N2}) = 4.0$\times$10$^{-29}$ & \\
  & k$_0$(\ce{CO2}) = 4.6$\times$10$^{-29}$ & \\
   & k$_0$(\ce{H2}) = 8.0$\times$10$^{-29}$ & \\
\hline
\end{tabular}
\end{table*}

Different reactions tend to dominate in different regions of an atmosphere. In the diffuse upper atmosphere (thermosphere), incoming UV radiation breaks apart dominant atmospheric species to produce radicals. In the dense lower atmosphere (troposphere), radicals can be transported from the upper atmosphere via turbulent mixing, or produced by lightning and/or GCRs. In this lower region, there is also sufficient pressure to collisionally deexcite the vibrationally excited intermediates in three-body reactions.

One newly discovered reaction with a great potential to produce substantial amounts \ce{H2CO} in upper atmospheres is \ce{^1O + CH3 -> H2CO + H}. Firstly, there will likely be high concentrations of reactants \ce{^1O} and \ce{CH3} in the upper atmospheres of planets containing \ce{CO2} and \ce{CH4}, as the former are the direct photodissociation fragments of the latter. Secondly, this reaction has a barrierless rate coefficient of $k$(298 K) = 4.3$\times$10$^{-10}$ cm$^{3}$s$^{-1}$, which is in the 94th percentile for highest two-body reaction rate coefficients in this study. For these reasons, we expect this reaction to be a dominant source of \ce{H2CO} in \ce{CO2}-rich and \ce{CH4}-containing atmospheres such as the early Earth. At the CCSD/aug-cc-pVDZ level of theory, we calculate this rate coefficient to be only 14\% lower (3.7 $\times$10$^{-10}$ cm$^{3}$s$^{-1}$), suggesting this calculation is not very sensitive to the choice of computational method.

In lower planetary atmospheres, we find two new three-body reactions that may be important pathways to \ce{H2CO}. These reactions are \ce{^1O + ^1CH2 + M -> H2CO + M} and \ce{^3O + ^3CH2 + M -> H2CO + M}. These reactions are most favourable at the high-pressure limit, where their rate coefficients are $k_{\infty}$(298 K) = 3.3$\times$10$^{-10}$ and 6.7$\times$10$^{-11}$ cm$^{3}$s$^{-1}$, respectively. The pressures at which these reaction rate coefficients reach 90\% of $k_{\infty}$(298 K) in a \ce{N2} bath gas are 0.61 bar and 7.1 bar, respectively. Such pressures would have been present in the evolving early Earth atmosphere $\sim$4.5 billion years ago \citep{2007SSRv..129...35Z}.

For new potentially important routes to HCN, we find \ce{^1O + H2CN -> HCN + OH}, which has a rate coefficient of $k$(298 K) = 1.2$\times$10$^{-13}$ cm$^{3}$s$^{-1}$. This reaction has the potential to be an important source of HCN in upper atmospheres with high \ce{CO2} mixing ratios, and low \ce{H2} and \ce{CH4} mixing ratios. The reason for this is that there is a direct competing reaction for HCN production from  \ce{H2CN + H -> HCN + H2}, which has a rate coefficient of $k$(298 K) = 2.2$\times$10$^{-11}$ cm$^{3}$s$^{-1}$. Therefore, the \ce{^1O}/H ratio in upper atmospheres will determine which of these two reactions dominates. We note also that this reaction has a complex reaction scheme, with two other favourable channels from the \ce{H2CNO*} intermediate: \ce{HCNO + H} and \ce{^3O + H2CN}. Our calculations of this reaction rate coefficient using two other computational methods ($\omega$B97XD, CCSD) suggests the channel to \ce{HCN + OH} may be more favourable than our BHandHLYP calculation implies, by up to a factor of $\sim$700 (see theoretical case study 9 in the SI for more details). Given these discrepancies, and the novelty of this reaction, we recommend experimental measurements be performed for the three product channels of \ce{^1O + H2CN}.

A new reaction with a great potential to destroy \ce{H2CO} is \ce{H2CO + ^1O -> HCO + OH}, which has a barrierless rate coefficient of 4.6$\times$10$^{-10}$ cm$^{3}$s$^{-1}$ at 298 K. As with the main new production pathway to \ce{H2CO}, this rate coefficient is one of the highest two-body rate coefficients in this study, and likely plays a role of attenuating \ce{H2CO} in upper atmospheres. At the CCSD/aug-cc-pVDZ level of theory, we calculate this rate coefficient to be only 50\% lower (2.3$\times$10$^{-10}$ cm$^{3}$s$^{-1}$) than the value at the BHandHLYP/aug-cc-pVDZ level of theory.

Lastly, we highlight a new HCN destruction pathway in lower atmospheres, \ce{^1O + HCN + M -> HCNO + M}. This reaction may be particularly important in attenuating HCN abundances in the troposphere, which is the region where HCN dissolves in rain droplets and makes its way into surface water. This reaction rate coefficient reaches 90\% of $k_{\infty}$(298 K) in a \ce{N2} bath gas at 3 bar.

\subsection*{CRAHCN-O}

CRAHCN-O is a chemical reaction network that can be used to simulate the production of HCN and \ce{H2CO} in atmospheres ranging from $\sim$50--400 K dominated by any of the following gases: \ce{CO2}, \ce{N2}, \ce{H2O}, \ce{CH4}, and \ce{H2}. CRAHCN-O is the amalgamation of the CRAHCN network developed in \citet{Pearce2020a} and the oxygen extension developed in this work. CRAHCN-O contains experimental rate coefficients (when available), and our consistently calculated theoretical rate coefficients from this work otherwise.

We summarize the oxygen extension in Tables S1 and S2 in the supplementary materials. In addition to the 126 reactions explored in this work, we include one experimental spin-forbidden collisionally induced intersystem crossing reaction (\ce{^1O + M -> ^3O + M}), whose rate coefficient cannot be calculated using our theoretical method. 

The original CRAHCN network can be found in the appendices of \citet{Pearce2020a}.

\section*{Conclusions}\label{conclusions}

In this work, we use a novel technique making use of computational quantum chemistry and experimental data to build a consistent reduced atmospheric hybrid chemical network oxygen extension (CRAHCN-O). This network can be used to simulate HCN and \ce{H2CO} chemistry in planetary atmospheres dominated by \ce{CO2}, \ce{N2}, \ce{H2O}, \ce{CH4}, and \ce{H2}.

The oxygen extension contains 127 reactions, and is made up of approximately 30\% experimental and 70\% consistently calculated theoretical rate coefficients. Below are the main conclusions of this work in bullet point.

\begin{itemize}
\item We discover 45 previously unknown reactions, and are the first to calculate their rate coefficients. These new reactions typically involve electronically excited species (e.g., \ce{^1O, ^1CH2, ^2N}).
\item The majority ($\sim$62\%) of our calculated rate coefficients are accurate to within a factor of two of experimental measurements. $\sim$84\% are accurate to within a factor of 6 of experimental values, and the rest are accurate to within about an order of magnitude of experimental values. This level of accuracy is consistent with the uncertainties assigned in large scale experimental data evaluations.
\item We identify 6 potentially key new production and destruction pathways for \ce{H2CO} and \ce{HCN} from these previously unknown reactions. 
\item The high, barrierless rate coefficient of \ce{^1O + CH3 -> H2CO + H} ($k$(298 K) = 4.3$\times$10$^{-10}$ cm$^{3}$s$^{-1}$) likely makes it a key source of formaldehyde in upper atmospheres where \ce{^1O} and \ce{CH3} are produced from the UV photodissociation of \ce{CO2} and \ce{CH4}, respectively.
\item Conversely, the high, barrierless rate coefficient of \ce{H2CO + ^1O -> HCO + OH} ($k$(298 K) = 4.6 $\times$10$^{-10}$ cm$^{3}$s$^{-1}$) likely makes it a key sink for formaldehyde in upper atmospheres.
\item \ce{^1O + H2CN -> HCN + OH} is less efficient than the known HCN source, \ce{H2CN + H -> HCN + H2}; However the former may dominate HCN production in \ce{CO2}-rich upper atmospheres with high \ce{^1O}/H ratios from \ce{CO2} photodissociation. \\
\item In lower atmospheres (i.e. high partial pressures), \ce{H2CO} may form via new reactions between \ce{^1O + ^1CH2} and \ce{^3O + ^3CH2}, which require a collisional third body at the high pressures present in these regions. HCN may be efficiently removed in this region via \ce{^1O + HCN + M -> HCNO + M}. 
\end{itemize}

Having now filled in the missing chemical data relevant to HCN and \ce{H2CO} production in \ce{CO2}- and \ce{H2O}-rich atmospheres, we intend to couple CRAHCN-O to a 1D chemical kinetic model to simulate the atmosphere of the early Earth.

\section*{Supporting Information}

Rate coefficient data, experimental data, Lennard-Jones parameters, theoretical case studies, and quantum chemistry data.

\section*{Acknowledgments}

B.K.D.P. is supported by an NSERC Alexander Graham Bell Canada Graduate Scholarship-Doctoral (CGS-D). P.W.A is supported by NSERC, the Canada Research Chairs, and Canarie. R.E.P. is supported by an NSERC Discovery Grant. We acknowledge Compute Canada for allocating the computer time required for this research.

\section*{References}

\bibliography{Bibliography_Rates}

\beginsupplement

\section*{Supporting Information}

\subsection*{CRAHCN-O}

In Tables~\ref{CRAHCN1} and \ref{CRAHCN2}, we display the Lindemann and the Arrhenius coefficients for the new oxygen reactions in CRAHCN-O. These rate coefficients consist of experimental values when available, and our consistently calculated theoretical values otherwise.

\setlength\LTcapwidth{\textwidth}
\begin{longtable*}{clccc}
\caption{Lindemann coefficients for the three-body oxygen reactions in the consistent reduced atmospheric hybrid chemical network oxygen extension (CRAHCN-O), valid within the 50--400 K temperature range. Experimental values are used when available, and calculated rate coefficients from this work are used otherwise. $k_{\infty}$ is the second-order rate coefficient in the high pressure limit with units cm$^{3}$s$^{-1}$. $k_0$ is the third-order rate coefficient in the low pressure limit with units cm$^{6}$s$^{-1}$. These values fit into the pressure-dependent rate coefficient equation $k$ = $\frac{k_0 [M] / k_{\infty}}{1 + k_0 [M] / k_{\infty}} k_{\infty}$. \label{CRAHCN1}} \\
No. & Reaction equation & k$_\infty$(298) &  k$_0$(298) & Source(s) \\ \hline \\[-2mm]
1. & \ce{CO2 + ^1O + M -> CO3 + M} & 1.6$\times$10$^{-10}$ & (M=\ce{N2}) 3.0$\times$10$^{-29}$ & Avg. of exper. vals, This work \\
 & & & (\ce{CO2}) 3.1$\times$10$^{-29}$ & This work \\
  & & & (\ce{H2}) 6.7$\times$10$^{-29}$ & This work \\
2. & \ce{HCO + ^2N + M -> HCON* + M* ->} & 2.0$\times$10$^{-11}$ & (\ce{N2}) 5.0$\times$10$^{-30}$ & This work \\
& \ce{HCNO + M} & & (\ce{CO2}) 5.6$\times$10$^{-30}$ & This work \\
\hline
\end{longtable*}

\setlength\LTcapwidth{\textwidth}
\begin{longtable*}{clccccc}
\caption{Arrhenius coefficients for the one- and two-body oxygen reactions in the consistent reduced atmospheric hybrid chemical network oxygen extension (CRAHCN-O). Experimental values are used when available, and calculated rate coefficients from this work are used otherwise. For the reactions with barriers from this work, rate coefficients are calculated at 50, 100, 200, 298, and 400 K, and are fit to the modified Arrhenius expression $k(T) = \alpha \left(\frac{T}{300}\right)^{\beta} e^{-\gamma/T}$. Barrierless reaction rate coefficients do not typically vary by more than a factor of 1--3 for temperatures between 50 and 400 K \citep{Reference587,Reference495,Reference481,Reference577,Reference572}.   Intermediate molecules are labelled with a bullet, and are included to describe the precise reaction pathway for multi-step reactions. First- and second-order reactions with rate coefficients slower than $k$(298 K) = 10$^{-21}$ are not included in this network. First-order rate coefficients have units s$^{-1}$. Second-order rate coefficients have units cm$^{3}$s$^{-1}$. \label{CRAHCN2}} \\
No. & Reaction equation & Forw./Rev. & $\alpha$ &  $\beta$ & $\gamma$ & Source\\ \hline \\[-2mm]
32. & \ce{NCCO -> CO + CN} & F & 3.4$\times$10$^{+7}$ & 14.33 & 12716 & This work\\
33. & \ce{CO2 + ^1O -> ^1CO3* -> ^3CO3* ->} & F & 7.4$\times$10$^{-11}$ & 0 & -133 & \citet{Reference1084}\\
 & \ce{CO2 + ^3O} & & & & & \\
34. & \ce{CO2 + ^2N -> NCO2* -> OCNO* ->} & F & 4.6$\times$10$^{-13}$ & 0 & 0 & Avg. of exper. vals  \\
 & \ce{CO + NO} & & & & & \\
\hline
\end{longtable*}

\subsection*{Experimental Data}

Experiments and reviews have measured and suggested reaction rate coefficients for several of the reactions in this network at or near $\sim$298 K. These values are listed in Table~\ref{TableExp}.

\setlength\LTcapwidth{\textwidth}
\begin{longtable*}{ccccc}
\caption{All available experimental or recommended reaction rate coefficients for the reactions in this study. For brevity, only the 10 most recent measurements of \ce{OH + CH4 -> H2O + CH3}, \ce{OH + H2 -> H2O + H}, and \ce{^3O + CH3 -> H2CO + H} are included; for a complete listing, we refer the
reader to the NIST Chemical Kinetics Database \citep{Reference494}. First-order rate coefficients have units s$^{-1}$. Second-order rate coefficients have units cm$^{3}$s$^{-1}$. Third-order rate coefficients have units cm$^{6}$s$^{-1}$.\label{TableExp}} \\
k(298K) & Technique & Temp. (K) & Pressure (Torr) & Reference(s)\\ \hline \\[-2mm]
\ce{CO2 + ^1O -> products} & & & & \\
2.3$\times$10$^{-10}$ & M & 300 & 20 & \citet{Reference1008} \\
1.4$\times$10$^{-10}$ & M & 295 & 25--250 & \citet{Reference1096} \\
1.0$\times$10$^{-10}$ & M & 297 & 40 & \citet{Reference1103} \\
$>$1.0$\times$10$^{-12}$ & M & & & \citet{Reference1102} \\
\end{longtable*}

\subsection*{Lennard-Jones Parameters}

\setlength\LTcapwidth{\textwidth}
\begin{longtable*}{lccccc}
\caption{Lennard-Jones force constants used in this study. Values are obtained from viscosity data when possible. \label{TableLJ}} \\
Molecule & $\sigma$ ($\AA$) & $\epsilon/k_b$ (K) & Source \\ \hline \\[-2mm]
\ce{CO3} & $^a$3.996 & $^a$190 & \citet{Reference2056} \\
\ce{NCCO} & $^b$4.38 & $^b$339 & \citet{Reference2056} \\
\ce{CO2} & 3.996 & 190 & \citet{Reference2056} \\
\ce{CH3CHO} & 3.97 & 436 & \citet{jetsurf2.0} \\
\ce{H2CNOH} & $^e$3.585 & $^e$507 & \citet{Reference2056} \\
\end{longtable*}

\subsection*{Theoretical Case Studies}

The following case studies provide additional details for some of the non-standard reactions in this study. Examples include intersystem crossing reactions, reactions with vibrational intermediates or complex pathways, and reactions where BHandHLYP/aug-cc-pVDZ misdiagnosed the barrier.

\subsubsection*{Case Study 1: \\ \ce{CO2 + ^1O -> ^1CO3* -> ^3CO3* -> CO2 + ^3O}}

The deexcitation of \ce{^1O} by \ce{CO2} has been studied considerably both experimentally and theoretically \citep{Reference1004,Reference1003,Reference1005,Reference1000,Reference1011,Reference1002,Reference1001,Reference1006,Reference1010,Reference1007,Reference1009,Reference1008}.

Experiments confirm that the dominant quenching pathway leads to ground state oxygen atoms (\ce{^3O}) \citep{Reference1012,Reference1007,Reference1009,Reference1008,Reference1001}. RRKM and statistical models have been used to explore the quenching mechanism \citep{Reference1003,Reference1006}, the dominant of which is to react \ce{^1O} by \ce{CO2} to form singlet \ce{CO3}, which then undergoes intersystem crossing to the triplet \ce{CO3} potential energy surface before decaying into \ce{^3O + CO2}.

An experiment by \citet{Reference1000} measures the singlet PES quenching pathway, \ce{CO2 + ^1O -> ^1CO3* -> CO + O2}, to be approximately 1000 times less efficient than the dominant mechanism.

Experimentally measured rate coefficients for the overall quenching of \ce{^1O} by \ce{CO2} in the 295--300 K range from 1.0--2.3$\times$10$^{-10}$ cm$^{3}$s$^{-1}$ \citep{Reference1002,Reference1008,Reference1011,Reference1012,Reference1084,Reference1096,Reference1103}.

In this work, we calculate the rate coefficient for \ce{CO2 + ^1O -> CO3} at the BHandHLYP/aug-cc-pVDZ level of theory to be$k$(298 K) =  3.8$\times$10$^{-11}$ cm$^{3}$s$^{-1}$, and assume this to be the rate-limiting step in the quenching pathway to \ce{CO2 + ^3O}. This value is only a factor of 3 lower than the nearest experimental measurement by \citet{Reference1103}.

We also include the third-order reaction \ce{CO2 + ^1O + M -> CO3 + M} in our network for M = \ce{N2}, \ce{CO2}, and \ce{H2}.

\subsubsection*{Case Study 2: \\ \ce{CO2 + ^2N -> NCO2* -> OCNO* -> CO + NO}}

Experimental measurements of the rate coefficient for this reaction at 300 K are between 1.8--6.8$\times$10$^{-13}$ cm$^{3}$s$^{-1}$ \citep{Reference554,Reference555,Reference557,Reference559,Reference1013}.
\citet{Reference549} reviewed these experiments and suggested a value of 3.6$\times$10$^{-13}$ cm$^{3}$s$^{-1}$.

There have been no theoretical studies performed on this reaction to date.

\citet{Reference557} suggest this reaction has a small energy barrier due to its fairly slow rate coefficient for a reaction of high exothermicity.

We do not find a barrier for this reaction at the BHandHLYP/aug-cc-pVDZ level of theory. We also do not find a barrier at the CCSD/aug-cc-pVDZ level of theory.

On the other hand, at the HF/aug-cc-pVDZ level of theory, we find a barrier of 34.3 kJ mol$^{-1}$ at the transition state for a C-N bond distance of 1.89$\AA$.

In a computational methods comparison study on the reaction of \ce{CH4 + H -> CH3 + H2}, we found the variational transition state barrier at the HF/aug-cc-pVDZ level of theory to be approximately twice the size of the barrier calculated at the BHandHLYP/aug-cc-pVDZ level of theory. We insert an artificial barrier of half the HF value (17.15 kJ mol$^{-1}$) into the calculation for $k$(298 K) at the BHandHLYP/aug-cc-pVDZ level of theory, and obtain a rate coefficient of 3.2$\times$10$^{-14}$ cm$^{3}$s$^{-1}$. This value is $\sim$6 times smaller than the nearest experimental value.

\subsubsection*{Case Study 3: \\ \ce{CO2 + CH -> CHCO2* -> HCOCO* -> HCO + CO}}

The rate coefficients for this reaction have been measured experimentally at 298 K and range from 1.8--2.1$\times$10$^{-12}$ cm$^{3}$s$^{-1}$ \citep{Reference1015,Reference1016,Reference1017,Reference1018}.
\citet{Reference451} reviewed the earliest of these experimental results and has suggested a $k$(298 K) value of 1.8$\times$10$^{-12}$ cm$^{3}$s$^{-1}$. \citet{Reference1018} predict this reaction to have little of no activation barrier below 400 K.

We find no theoretical studies of this reaction.

We find the first step of this reaction to be the rate-limiting step. At the BHandHLYP/aug-cc-pVDZ level of theory, the first step of this reaction has a barrier; However, at the B3LYP/aug-cc-pVDZ level of theory, this reaction step is barrierless. We remove the barrier from our calculation to match expectation from experiment \citep{Reference1018} and obtain an overall rate coefficient of $k$(298 K) = 3.1$\times$10$^{-12}$ cm$^{3}$s$^{-1}$ at the BHandHLYP/aug-cc-pVDZ level of theory. This is within a factor of 1.5 of the nearest experimental value.

\subsubsection*{Case Study 4: \ce{CO2 + ^3CH2 -> H2CO + CO}}

\citet{Reference1014} experimentally measured the rate coefficient for this reaction at 298 K to be 3.9$\times$10$^{-14}$ cm$^{3}$s$^{-1}$. \citet{Reference545} performed upper bound experiments on this reaction and found $k$(298) to be no greater than 1.4$\times$10$^{-14}$ cm$^{3}$s$^{-1}$.

\citet{Reference1192} studied this reaction theoretically, and found the lowest energy path is to form the \ce{^3CH2$\cdots$CO2} complex, followed the subsequent reaction into \ce{^3CH2CO2} over a 19.3 kcal mol$^{-1}$ barrier (at the G2 level of theory). They find that the lowest energy path from the \ce{^3CH2$\cdots$CO2} complex is to fragment back into \ce{^3CH2} and \ce{CO2} over a 1.1 kcal mol$^{-1}$ barrier. They also suggest an intersystem crossing reaction from this complex to the singlet surface is unlikely, and that \ce{^3CH2} would may require collisional reaction to the singlet state in order for this reaction to proceed.

We find similar results to \citet{Reference1192} at the BHandHLYP/aug-cc-pVDZ and CCSD/aug-cc-pVDZ levels of theory. \ce{CO2 + ^3CH2} proceeds via a barrierless reaction to form the \ce{^3CH2$\cdots$CO2} complex with a C-C bond distance of 3.25 (3.23) $\AA$ at the BHandHLYP/aug-cc-pVDZ (CCSD/aug-cc-pVDZ) level of theory. This reaction efficiently decays back into \ce{CO2 + ^3CH2}, and the barrier to \ce{^3CH2CO2} is 15.7 (19.3) kcal mol$^{-1}$ at the BHandHLYP/aug-cc-pVDZ (CCSD/aug-cc-pVDZ) level of theory. We find the rate coefficient for the reaction \ce{CO2 + ^3CH2 -> ^3CH2$\cdots$CO2 -> ^3CH2CO2} to be $k$(298 K) = 1.8$\times$10$^{-23}$ (2.3$\times$10$^{-25}$) cm$^{3}$s$^{-1}$ at the BHandHLYP/aug-cc-pVDZ (CCSD/aug-cc-pVDZ) level of theory, which is too slow to consider in our network.

Our results and those of \citet{Reference1192} suggest this reaction  likely does not occur on the triplet surface. For this reason, we do not include it in our network, and instead only include the singlet surface reaction \ce{CO2 + ^1CH2 -> H2CO + CO}. We find \ce{CO2 + ^1CH2 -> H2CO + CO} to have a rate coefficient of 8.0$\times$10$^{-13}$ cm$^{3}$s$^{-1}$ at 298 K, which is only a factor of 2 larger than the experimental value for \ce{CO2 + ^3CH2 -> H2CO + CO}. This adds some evidence to the suggestion that \ce{^3CH2} must first collisionally excite to \ce{^1CH2} before reacting with \ce{CO2} to produce \ce{H2CO + CO}.

\subsubsection*{Case Study 5: \ce{HCO + HCO -> products}}

Three potential product channels of the self-reaction of HCO have been reported experimentally \citep{Reference1033,Reference1034,Reference1035,Reference1036,Reference1037,Reference1038}.

\begin{equation}
\ce{HCO + HCO -> cis-C2H2O2}
\end{equation}
\begin{equation}
\ce{HCO + HCO -> H2CO + CO}
\end{equation}
\begin{equation}
\ce{HCO + HCO -> 2CO + H2}
\end{equation}

Rate coefficients for \ce{HCO + HCO -> H2CO + CO} have been experimentally measured at 295--298 K and range from 3.0--7.5$\times$10$^{-11}$ cm$^{3}$s$^{-1}$ \citep{Reference1034,Reference1035,Reference1036,Reference1037,Reference1092}..
The lack of temperature dependence in the range of 298--475 K suggests this reaction is barrierless \citep{Reference1036}.

\citet{Reference1033} performed the only experimental measurement of the rate coefficient for \ce{HCO + HCO -> 2CO + H2} at 298 K, which they find to be 3.6 $\times$10$^{-11}$ cm$^{3}$s$^{-1}$.

Rate coefficients for \ce{HCO + HCO -> trans-C2H2O2} have been experimentally measured at 298 K to be in the range of 2.8--500$\times$10$^{-13}$ cm$^{3}$s$^{-1}$ \citep{Reference1033,Reference1109}.

There are no experimental measurements of the rate coefficient for \ce{HCO + HCO -> cis-C2H2O2}.

\citet{Reference1039} performed theoretical quantum computational simulations, and found the most important product channels to be:

\begin{equation}
\ce{HCO + HCO -> cis-C2H2O2* -> 2CO + H2}
\end{equation}
\begin{equation}
\ce{HCO + HCO -> trans-C2H2O2* -> anti-HCOH + CO}
\end{equation}


We find no direct abstraction reaction for \ce{HCO + HCO -> H2CO + CO} on the singlet surface. We do however find an inefficient abstraction reaction for \ce{HCO + HCO -> anti-HCOH + CO}, with a $k$(298 K) rate coefficient of 3.4$\times$10$^{-24}$ cm$^{3}$s$^{-1}$ at the BHandHLYP/aug-cc-pVDZ level of theory.

We find the reaction \ce{HCO + HCO -> trans-C2H2O2} to have a barrierless rate coefficient of 4.1$\times$10$^{-13}$ cm$^{3}$s$^{-1}$ at 298 K at the BHandHLYP/aug-cc-pVDZ level of theory. At the same level of theory, we calculate the rate coefficient of the subsequent reaction \ce{trans-C2H2O2 -> anti-HCOH + CO} (k(298 K) = 7.1$\times$10$^{-34}$ s$^{-1}$) to be slightly smaller than than the decay back into \ce{HCO + HCO} (k(298 K) = 1.7$\times$10$^{-33}$ s$^{-1}$). Finally, we find \ce{anti-HCOH} efficiently isomerizes into \ce{H2CO}. We calculate the overall rate coefficient at 298 K for \ce{HCO + HCO -> trans-C2H2O2 -> anti-HCOH + CO -> H2CO + CO} to be 1.2$\times$10$^{-13}$ s$^{-1}$, which is slightly reduced from the barrierless value due to the slight preference in the decay back to \ce{HCO + HCO} over \ce{anti-HCOH + CO}. This is only a factor of 2 smaller than the nearest experimental value for \ce{HCO + HCO -> trans-C2H2O2} \citep{Reference1033}.

We calculate the reaction \ce{HCO + HCO -> cis-C2H2O2* -> CO + CO + H2} to have a rate coefficient of $k$(298 K) = 7.4 $\times$10$^{-11}$ cm$^{3}$s$^{-1}$ at the BHandHLYP/aug-cc-pVDZ level of theory. This is only a factor of 2 larger than the only experimental value.

\subsubsection*{Case Study 6: \ce{CO + OH -> intermediates* -> CO2 + H}}

Experimental measurements of the rate coefficient for this reaction at 296--300 K range from 8.5$\times$10$^{-14}$ to 9.7$\times$10$^{-13}$ cm$^{3}$s$^{-1}$ \citep{Reference1163,Reference1165,Reference1166,Reference1167,Reference1168,Reference1169,Reference1170,Reference1171,Reference1172,Reference1173}.
\citet{Reference451} review the kinetic data from experiments and suggest a very slight temperature dependence, suggesting this reaction proceeds with little or no reaction barrier.

There are multiple reaction pathways for this reaction, but the fastest is that which proceeds through the \ce{OH$\cdots$CO*} and \ce{cis-HOCO*} intermediates \citep{Reference1175}.

At the BHandHLYP/aug-cc-pVDZ level of theory, we calculate $k$(298 K) for the barrierless first step, \ce{CO + OH -> OH$\cdots$CO*}, to be 9.7$\times$10$^{-12}$ cm$^{3}$s$^{-1}$. However, we find intermediate barriers at the second and third steps of this calculation at the BHandHLYP/aug-cc-pVDZ level of theory, making the overall rate coefficient for \ce{CO + OH -> OH$\cdots$CO* -> cis-HOCO* -> CO2 + H}, 5.7$\times$10$^{-19}$ cm$^{3}$s$^{-1}$. This is several orders of magnitude smaller than the range of experimental values. At the B3LYP/aug-cc-pVDZ level of theory, these barriers are more comparable, resulting in only a factor of 3.4 reduction between the barrierless first step and the overall rate coefficient. {\it Ab initio} calculations show similar barrier heights to our the B3LYP/aug-cc-pVDZ calculations \citep{Reference1175}.

We reduce the calculated rate coefficient for the barrierless first step at the BHandHLYP/aug-cc-pVDZ level of theory by a factor of 3.4 to match the barrier effects at the B3LYP/aug-cc-pVDZ level of theory. This gives us a rate coefficient of $k$(298) = 2.9$\times$10$^{-12}$ cm$^{3}$s$^{-1}$, which is a factor of 3 higher than the nearest experimental value.

\subsubsection*{Case Study 7: \ce{OH + O -> HO2$_{(\nu)}$* -> O2 + H}}

Experimental measurements for this reaction at 298--300 K range from 2.3--4.3$\times$10$^{-11}$ cm$^{3}$s$^{-1}$ \citep{Reference1137,Reference1138,Reference1139,Reference1140,Reference1141,Reference1142,Reference1143,Reference1144,Reference1145,Reference1146}.

This reaction proceeds through \ce{HO2}, which, in its ground vibrational state has been noted to be long-lived \citep{Reference1147,Reference1148}. Our calculations confirm that the decay of \ce{HO2} into \ce{O2 + H} is slow ($<$ 10$^{-47}$ s$^{-1}$). This suggests this reaction proceeds through an excited vibrational state, as is to be expected when two reactants combine to form a single product \citep{Vallance_Book}.

We calculate the rate coefficient of \ce{OH + ^3O -> HO2} at 298 K at the BHandHLYP/aug-cc-pVDZ level of theory to be 7.4$\times$10$^{-11}$ cm$^{3}$s$^{-1}$, and assume the subsequent vibrational decay into \ce{O2 + H}. Our calculated rate coefficient is within a factor of 2 of the experimental range.

There are currently no experimental measurements for the rate coefficient of \ce{OH + ^1O -> O2 + H}, which we find also proceeds through \ce{HO2}. We calculate the rate coefficient of \ce{OH + ^1O -> HO2} at 298 K at the BHandHLYP/aug-cc-pVDZ level of theory to be 1.0$\times$10$^{-9}$ cm$^{3}$s$^{-1}$, and similarly assume the vibrational decay into \ce{O2 + H}.

\subsubsection*{Case Study 8: \ce{OH + NH -> products}}

No experiments have been performed to date on the reaction of \ce{OH + NH}. \citet{Reference1053} use analogous reactions to suggest rate coefficients of $k$(298 K) = 3.3$\times$10$^{-11}$ cm$^{3}$s$^{-1}$ and 5.1$\times$10$^{-12}$ cm$^{3}$s$^{-1}$ for \ce{OH + NH -> HNO + H} and \ce{OH + NH -> H2O + ^4N}, respectively. They suggest little or no barrier exists for either pathway.

\citep{Klippenstein2009} performed theoretical transition state theory calculations for this reaction using a range of computational quantum methods. They calculated 298 K reaction rate coefficients of 6.8$\times$10$^{-11}$ cm$^{3}$s$^{-1}$ and 1.4$\times$10$^{-12}$ cm$^{3}$s$^{-1}$ for \ce{OH + NH -> HNO + H} and \ce{OH + NH -> H2O + ^4N}, respectively.

We find the \ce{OH + NH -> HNO + H} reaction to proceed through multiple intermediates, including \ce{OH$\cdots$NH*}, \ce{trans-HNOH*}, and \ce{H2NO*}. We calculate the barrierless first step of this reaction \ce{OH + NH -> OH$\cdots$NH*} at 298 K at the BHandHLYP/aug-cc-pVDZ level of theory to be 7.0$\times$10$^{-12}$ cm$^{3}$s$^{-1}$. However, at this level of theory, we find a large forward barrier at the third reaction step (i.e., \ce{trans-HNOH* -> HNO + H}), which reduces the overall rate coefficient to 2.6$\times$10$^{-14}$ cm$^{3}$s$^{-1}$. This is over three orders of magnitude smaller than the recommended and theoretical values. Conversely, at the B3LYP/aug-cc-pVDZ level of theory, the third forward reaction step barrier is smaller than the reverse barrier, which makes the barrierless first step the rate limiting step.

We remove the barrier at the third reaction step from our calculation to match the kinetic data and theoretical studies, and obtain an overall rate coefficient for \ce{OH + NH -> HNO + H} of $k$(298 K) = 7.0$\times$10$^{-12}$ cm$^{3}$s$^{-1}$ at the BHandHLYP/aug-cc-pVDZ level of theory. This value is a factor of 5 smaller than the suggested value \citep{Reference1053}.

We calculate the rate coefficient for \ce{OH + NH -> H2O + ^4N} at 298 K at the BHandHLYP/aug-cc-pVDZ level of theory to be 6.8$\times$10$^{-13}$ cm$^{3}$s$^{-1}$. This is a factor of 5 smaller than the suggested value by \citet{Reference1053}.

\subsubsection*{Case Study 9: \ce{O + H2CN -> CH2NO* -> products}}

No experiments to date have measured the rate coefficient of \ce{^3O + H2CN -> HCN + OH}. \citet{Reference446} suggested a temperature-independent rate of 8.3$\times$10$^{-11}$ cm$^{3}$s$^{-1}$ based on calculations using published chemical compositions of the flames of methane, nitrogen and oxygen at $>$1850 K. They note that this calculation does not include the effects of an energy barrier, and thus this value is not reliable at 298 K. \citet{Reference446} also suggested this same rate coefficient for the reaction \ce{H + H2CN -> HCN + H2}.

No previous theoretical studies regarding this reaction have been performed.

\begin{figure}[!hbtp]
\centering
\includegraphics[width=\linewidth]{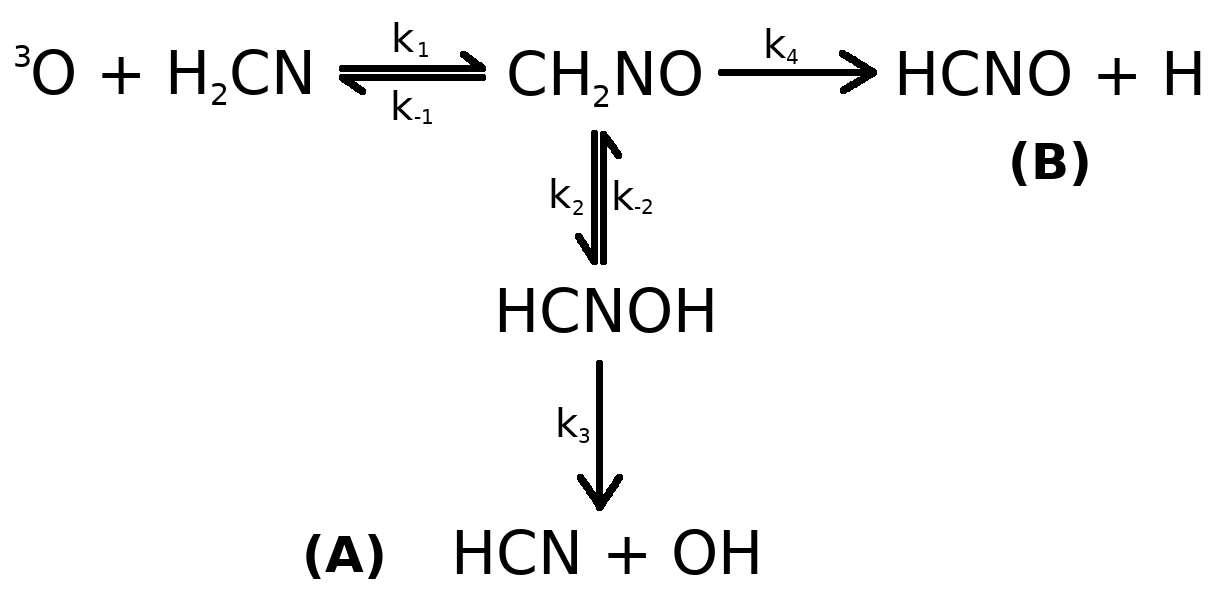}
\caption{Mechanistic model for the reaction of \ce{^3O + H2CN}. Two efficient product channels on the doublet surface exist: {\bf (A)} \ce{HCN + OH} and {\bf (B)} \ce{HCNO + H}.}
\label{MechModel}
\end{figure}

Unlike \ce{H + H2CN -> HCN + H2}, which we found in previous work to proceed efficiently through a barrierless abstraction mechanism \citep{Reference598}, we find no abstraction pathway for \ce{^3O + H2CN -> HCN + OH}. Instead, we find that \ce{^3O} and \ce{H2CN} efficiently react to form \ce{CH2NO*} with a rate coefficient of 3.1$\times$10$^{-11}$ cm$^{3}$s$^{-1}$. This product most commonly decays back into the original reactants; However, there are two other favourable pathways. The mechanistic model for these reactions is shown in Figure \ref{MechModel}.

Using the mechanistic model above, we use the steady-state solutions to the kinetic rate equations to calculate the overall rate coefficients for paths A and B. This is done by equating the kinetic rate equations for each species in the mechanistic model to zero (e.g. d[\ce{CH2NO}]/dt = 0 = $k_1$[\ce{^3O + H2CN}] + $k_{-2}$[\ce{HCNOH}] - ($k_{-1}$ + $k_2$)[\ce{CH2NO}]), and substituting these equations into the overall kinetic rate equations for products A and B from the initial reactants. This gives us the following rate coefficients for paths A and B:

\begin{equation}
k_A = \frac{k_1 k_3}{\alpha},
\end{equation}
\begin{equation}
\alpha = \frac{\left(k_{-1} + k_2 + k_4\right)\left(k_{-2} + k_3\right)}{k_2} - k_{-2}.
\end{equation}

\begin{equation}
k_B = \frac{k_1 k_4}{\beta},
\end{equation}
\begin{equation}
\beta = k_{-1} + k_2 + k_4 - \frac{k_{-2} k_2}{k_{-2} + k_3}.
\end{equation}

We calculate these rate coefficients at the BHandHLYP/aug-cc-pVDZ level of theory to be k$_A$ = 4.0$\times$10$^{-14}$ and k$_B$ = 8.3$\times$10$^{-15}$ cm$^{3}$s$^{-1}$ at 298 K, respectively.

We propose that the suggested barrierless rate coefficient for \ce{^3O + H2CN -> HCN + OH} by \citet{Reference446} is not an accurate estimate for this overall reaction at 298 K. In fact, the large barrier for \ce{CH2NO* -> HCNOH*} isomerization at 298 K plays a key role in decreasing this overall rate coefficient. We find the isomerization barrier to also have similar heights when using the B3LYP and CCSD computational methods.

We use similar mechanistic modeling to calculate the rate coefficients for the reactions of \ce{^1O + H2CN -> CH2NO* -> products}; However along with the two decay pathways above, there is an additional decay pathway to \ce{^3O + H2CN}. We are the first to calculate these three \ce{^1O + H2CN} reaction rate coefficients.

At the BHandHLYP/aug-cc-pVDZ level of theory, we calculate the rate coefficients for the reaction of \ce{^1O + H2CN} to products (A) \ce{HCN + OH}, (B) \ce{HCNO + H}, and (C) \ce{^3O + H2CN}, to be 1.2$\times$10$^{-13}$, 6.0$\times$10$^{-13}$, and 4.5$\times$10$^{-10}$ cm$^{3}$s$^{-1}$, respectively.

Given the potential importance of \ce{^1O + H2CN -> HCN + OH} to produce HCN in atmospheres, and the similar barrier heights to the three product channels, we also calculated these rate coefficients at the $\omega$B97XD/aug-cc-pVDZ and CCSD/aug-cc-pVDZ levels of theory. 

At the $\omega$B97XD/aug-cc-pVDZ level of theory, we calculate the rate coefficients for the reaction of \ce{^1O + H2CN} to products (A), (B), and (C), to be 5.3$\times$10$^{-11}$, 2.8$\times$10$^{-10}$, and 6.6$\times$10$^{-23}$ cm$^{3}$s$^{-1}$, respectively.

At the CCSD/aug-cc-pVDZ level of theory, we calculate the rate coefficients for the reaction of \ce{^1O + H2CN} to products (A), (B), and (C), to be 8.8$\times$10$^{-11}$, 1.9$\times$10$^{-11}$, and 2.2$\times$10$^{-10}$ cm$^{3}$s$^{-1}$. 

In the case of BHandHLYP and CCSD, the dominant channel for the reaction of \ce{^1O + H2CN} is (C). This is not the case for $\omega$B97XD, where channel (C) is negligible, and the dominant channel is (B). The rate coefficient for the potentially important HCN source, channel (A), varies by a factor of 733 across the three levels of theory. Given these discrepancies, we recommend these reactions be followed up with an experimental study.

\subsubsection*{Case Study 10: \ce{O + CH2 -> products}}

\ce{^3O + ^3CH2} combine to form a vibrationally excited \ce{H2CO} molecule \citep{Reference1074}. In high atmospheric pressures, this molecule can be collisionally deexcited in the reaction

\begin{equation*}
\ce{^3O + ^3CH2 + M -> H2CO + M}.
\end{equation*}

However, in upper atmospheres, where collisions are less frequent, the vibrationally excited \ce{H2CO} will dissociate via 2 equally favourable pathways \citep{Reference1074,Reference451,Reference509}

\begin{equation*}
\ce{^3O + ^3CH2 -> H2CO$_{(\nu)}$* -> CO + H + H}
\end{equation*}
\begin{equation*}
\ce{^3O + ^3CH2 -> H2CO$_{(\nu)}$* -> CO + H2}
\end{equation*}

Experimental measurements of the rate coefficient of \ce{^3O + ^3CH2 -> products} at 295--296 K are 1.3$\times$10$^{-10}$ cm$^{3}$s$^{-1}$ \citep{Reference1072,Reference1073}. Reviews of this reaction over a range of temperatures and pressures suggest a wider range of 1.9$\times$10$^{-11}$--2.0$\times$10$^{-10}$ cm$^{3}$s$^{-1}$ \citep{Reference509,Reference451}.

We calculate the rate coefficient of \ce{^3O + ^3CH2 -> H2CO} at the BHandHLYP/aug-cc-pVDZ level of theory to be $k$(298 K) = 6.7$\times$10$^{-11}$ cm$^{3}$s$^{-1}$, which is within the range of suggested values, and only a factor of 2 lower than the two experimental measurements. We allow this reaction to proceed along the two equally favourable dissociation channels, each with a calculated rate coefficient of 3.4$\times$10$^{-11}$ cm$^{3}$s$^{-1}$.

Excited oxygen (\ce{^1O}) and methylene (\ce{^1CH2}) also react to produce vibrationally excited \ce{H2CO}. 

We calculate the rate coefficient of \ce{^1O + ^1CH2 -> H2CO} at the BHandHLYP/aug-cc-pVDZ level of theory to be $k$(298 K) = 3.3$\times$10$^{-10}$ cm$^{3}$s$^{-1}$. We assume that the two dissociation pathways for vibrationally excited \ce{H2CO} are also equally favourable for this reaction, and allow this reaction to proceed to form \ce{CO + H + H} and \ce{CO + H2} with equal rate coefficients of 1.7$\times$10$^{-10}$ cm$^{3}$s$^{-1}$.

\subsubsection*{Case Study 11: \ce{^1O + CH4 -> CH3OH$_{(\nu)}$ -> OH + CH3}}

\ce{^1O} and \ce{CH4} mainly react to form vibrationally excited \ce{CH3OH}, the dominant subsequent pathway of which is to produce \ce{OH + CH3} \citep{Reference1100,Reference1101,Reference1080,Reference1095,Reference1096,Reference1097,Reference1098}.

Experimental measurements of \ce{^1O + CH4 -> OH + CH3} from 295--300 K range from 1.4--3.8$\times$10$^{-10}$ \citep{Reference1080,Reference1095,Reference1096,Reference1097,Reference1098,Reference1099}.

We calculate the rate coefficient of \ce{^1O + CH4 -> CH3OH} at 298 K with the BHandHLYP/aug-cc-pVDZ level of theory to be 5.8$\times$10$^{-9}$ cm$^{3}$s$^{-1}$, and assume the vibrational decay into \ce{OH + CH3} as suggested. Our calculated rate coefficient is a factor of 15 larger than the nearest experimental value.

\subsubsection*{Case Study 12: \ce{^1O + H2 -> H2O$_{(\nu)}$ -> OH + H}}

Experimental measurements of the rate coefficient for \ce{^1O + H2 -> OH + H} at 298--300 K are between 1.1 and 3.0$\times$10$^{-10}$ cm$^{3}$s$^{-1}$ \citep{Reference1079,Reference1080,Reference1081,Reference1082,Reference1083,Reference1098}.

This reaction is known to proceed through vibrationally excited \ce{H2O} in its ground electronic state \citep{Reference1093,Reference1094}.

We calculate the rate coefficient for \ce{^1O + H2 -> H2O} at 298 K at the BHandHLYP/aug-cc-pVDZ level of theory to be 7.1$\times$10$^{-10}$ cm$^{3}$s$^{-1}$, and assume vibrational decay into \ce{OH + H}, as suggested. This calculated value is a factor of 2 larger than the nearest experimental value.

\subsection*{Quantum Chemistry Data}

\setlength\LTcapwidth{\textwidth}
\begin{longtable*}{cclcccccc}
\caption{Quantum Chemistry Simulation Data at the BHandHLYP/aug-cc-pVDZ level of theory. Cartesian coordinates are in angstroms. Energies are in kJ mol$^{-1}$. E$_e$ is the electronic energy, ZPE is the zero point energy, and q$_x$ are the partition functions (t: translational, e: electronic, v:vibrational, r:rotational). \label{quantumdata}} \\
Reac. & Species & Geometry (Atom, X, Y, Z) & E$_e$ + ZPE & q$_t$/V (m$^{-3}$) & q$_e$ & q$_v$ & q$_r$ \\ \hline \\[-2mm]
1/33 & \ce{CO2} & O, 0.00000, 0.00000, 1.15274 & -494917.399028 & 2.82E+32 &	1 &	1.07E+00 &	5.23E+02 \\
& & O, 0.00000, 0.00000, -1.15274 & & & & & \\
& & C, 0.00000, 0.00000, 0.00000 & & & & & \\
 & \ce{^1O} &  O, 0.00000, 0.00000, 0.00000 & -196788.921964 & 6.19E+31 &	1 &	1.00E+00 &	1.00E+00 \\
 & TS & O, 1.15275, 1.18505, 0.00000 & -691707.204162 & 4.49E+32 &	1	& 1.07E+00 &	1.26E+05 \\
\hline
\end{longtable*}

\end{document}